\documentclass[fleqn,usenatbib]{mnras}

\usepackage{newtxtext,newtxmath}

\usepackage[T1]{fontenc}

\DeclareRobustCommand{\VAN}[3]{#2}
\let\VANthebibliography\thebibliography
\def\thebibliography{\DeclareRobustCommand{\VAN}[3]{##3}\VANthebibliography}

\usepackage{graphicx}	

\usepackage{amsmath}	
\usepackage{amssymb}	
\usepackage{CJK}
\usepackage{hyperref}
\usepackage{color}
\usepackage{comment}
\usepackage{mathrsfs}

\newcommand\St{\,{\rm St}}
\usepackage{soul}
\def\YWc#1{\textcolor{cyan}{YW: #1}}
\newcommand{\rdnote}[1]{{\color{red} [R.D.: #1]}}
\newcommand{\YXC}[1]{{\color{red} [YX.: #1]}}

\definecolor{mygray}{gray}{0.6}
\definecolor{emerald}{RGB}{0,155,155}
\newcommand{\hjc}[1]{\textcolor{emerald}{[HJ: \textit{\small #1}]}}

\newcommand{\hjrem}[1]{\textcolor{mygray}{\sout{#1}}}

\title[Distinguishing Disc Winds from Viscosity]{Distinguishing Magnetized Disc Winds from Turbulent Viscosity through Substructure Morphology in Planet-forming Discs}

\author[Wu, Chen, Jiang et al., 2023]{Yinhao Wu (吴寅昊),$^{1}$\thanks{These authors contributed equally to this work.}
Yi-Xian Chen (陈逸贤),$^{2}$\footnote[1]~ 
Haochang Jiang (蒋昊昌),$^{3,4}$\footnote[1]~
\newauthor
Ruobing Dong (董若冰),$^{5,6}$~
Enrique Mac\'{i}as,$^{3}$~
Min-Kai Lin (林明楷),$^{7,8}$~
\newauthor
Giovanni P. Rosotti,$^{9,1}$~
Vardan Elbakyan$^{1,10}$~
\\
\\
$^{1}$School of Physics and Astronomy, University of Leicester, Leicester LE1 7RH, UK \href{mailto:email@domain}{yw505@leicester.ac.uk} (YW)\\
$^{2}$Department of Astrophysical Sciences, Princeton University, USA \href{mailto:email@domain} {yc9993@princeton.edu} (YC)\\
$^{3}$European Southern Observatory, Karl-Schwarzschild-Str 2, 85748 Garching, Germany \href{mailto:email@domain}{hjiang@eso.org} (HJ)\\
$^{4}$Department of Astronomy, Tsinghua University, Haidian DS 100084, Beijing, China\\
$^{5}$Department of Physics and Astronomy, University of Victoria, Victoria, BC V8P 5C2, Canada\\
$^{6}$Kavli Institute for Astronomy and Astrophysics, Peking University, Beijing 100871, China\\
$^{7}$Institute of Astronomy and Astrophysics, Academia Sinica, Taipei 10617, Taiwan\\
$^{8}$Physics Division, National Center for Theoretical Sciences, Taipei 10617, Taiwan\\
$^{9}$Dipartimento di Fisica ``Aldo Pontremoli'', Universita degli Studi di Milano, via Celoria 16, Milano, 20133, Italy\\
$^{10}$Research Institute of Physics, Southern Federal University, Rostov-on-Don 344090, Russia
}

\date{Accepted XXX. Received YYY; in original form ZZZ}

\pubyear{2023}

\begin{document}
\begin{CJK*}{UTF8}{gbsn}  
\label{firstpage}
\pagerange{\pageref{firstpage}--\pageref{lastpage}}
\maketitle

\begin{abstract}
The traditional paradigm of viscosity-dominated evolution of protoplanetary discs
has been recently challenged by existence of magnetised disc winds. 
However, 
distinguishing wind-driven and turbulence-driven accretion 
through observations has been difficult.
In this study, 
we present a novel approach to identifying their separate contribution to angular momentum transport by studying the gap and ring morphology of 
planet-forming discs in the ALMA continuum. 
We model the gap-opening process of planets in discs with both viscous evolution 
and wind-driven accretion 
by 2D multi-fluid hydrodynamical simulations. 
Our results show that gap-opening planets in wind-driven accreting discs generate characteristic dust substructures that differ from those in purely viscous discs.
Specifically, 
we demonstrate that discs where wind-driven accretion dominates the production of substructures exhibit significant asymmetries.
Based on the diverse outputs of mock images in the ALMA continuum, 
we roughly divide the planet-induced features into four regimes (moderate-viscosity dominated, 
moderate-wind dominated, 
strong-wind dominated, inviscid).
The classification of these regimes sets up a potential method to 
constrain the strength of magnetised disc wind and viscosity 
based on the observed gap and ring morphology. 
We discuss the asymmetry feature in our mock images and 
its potential manifestation in ALMA observations.
\end{abstract}




\section{Introduction}\label{sec: intro}

The Atacama Large Millimeter/submillimeter Array (ALMA) has discovered a rich catalogue of diverse features in mm/sub-mm observation of protoplanetary discs (PPDs), 
including cavities, 
gaps, rings and vortices \citep[e.g.,][]{ALMA2015,long2018,andrews2018,2018HuangDSHARP2,lodato2019,Long2019,Andrews2020}. 
Among those substructures, the most ubiquitous features are the bright rings and dark gaps, 
a classical interpretation for which is the perturbation of gas and dust profiles by an embedded planet \citep[e.g.,][]{PaardekooperEtal2022}. 
Due to excitation and dissipation of density waves launched at resonances, 
planets carve out deficits in the surrounding gas density profiles \citep{Goldreich_Tremaine_1980,  Lin_Papaloizou1986a, Lin_Papaloizou_1993}. In turn, a pressure bump forms at the outer edge of the gas gap, which efficiently traps in-drifting dust particles that have limited coupling to gas \citep{paardekooper2006,Zhuetal2011}, 
leading to bright dust rings. 
With suitable parameters for disc viscosity, 
thickness as well as planet mass, 
one can reproduce a variety of dust emission features similar to what's discovered in the sub-mm ALMA catalogue \citep{Rosotti2016,2018ZhangDSHARP}.

While the turbulent viscosity parameter $\alpha_{\rm v}$ \citep{Shakura-Sunyaev73}, acting to maintain disc profile and prevent gap opening, 
have been inferred to be low in planet-forming regions of PPDs \citep{Flaherty2017,Lodato2017,Dullemond2018,Rosotti2020,DoiKataoka2021}, which 
corroborates 
non-ideal magnetohydrodynamics (MHD) simulations \citep{2013BaiStone}, 
recent theoretical works have proposed that magnetically-driven disc winds might transport the angular momentum of gas at a faster rate than disc turbulence, 
dominating the PPD accretion \citep{Armitage2013,Suzuki2016,Hasegawa2017,Lesur2021,Cui2021,Cui2022}.

However, 
constraining the disc wind level from the observational side is challenging. 
A few pioneer works have tried to distinguish the dominant accretion mechanism by studying the protoplanetary disc size evolution from a population view \citep{TrapmanEtal2020,LongEtal2022,TrapmanEtal2022a,ZagariaEtal2022,TociEtal2023}, 
while the results remain unconcluded.
On the other hand,
current observation results can indeed constrain the upper limit of the disc turbulence coefficient to be as low as $10^{-4}$ to $10^{-3}$ \citep{Flaherty2017,Flaherty2020,Miotello2022,Giovanni-2023}. If protoplanetary discs are indeed nearly inviscid \citep[e.g.,][]{Villenave2020}, this serves as strong evidence for a mechanism other than the gas viscosity to explain how discs accrete and evolve, 
with MHD winds currently being a promising alternative \citep{Lesur2021}. 
Wind-dominated accretion could reconcile the low turbulent velocities with a moderate accretion rate onto the host star.

\begin{figure*}
\centering
\includegraphics[width=0.99\linewidth]{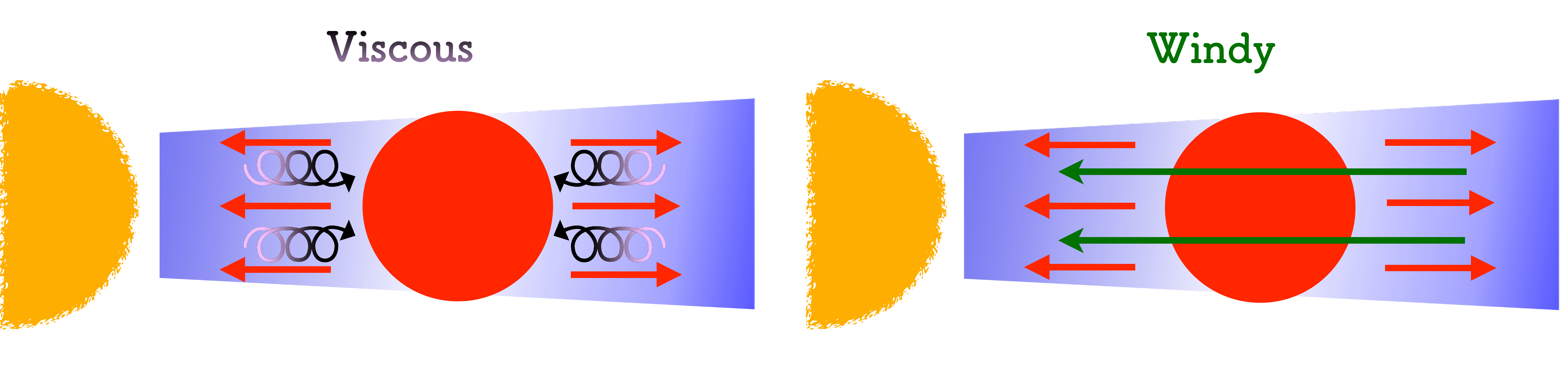}
\caption{An edge-on-view sketch illustrating the different components of dust/pebble transport around a gap in a planet-forming disc in the disc midplane region. 
Radial drift (shown in red), which results from the gas headwind and tailwind, leads to the formation of a dust gap/cavity in the disc. 
In viscous discs, pebble/dust diffusion (shown in blackpink) tends to fill the gap from both sides. 
In MHD windy discs, the wind-driven pebble flux, associated with the gas flux driven by wind torque (shown in green), is always directed toward the host star.}
\label{fig:cartoon}
\end{figure*}

Under this context, 
\citet{MHD-wind-Elbakyan} applied the torque formula for disc wind with an accretion parameter $\alpha_{\rm dw}$, 
introduced by \cite{Tabone22}, 
and investigated its effect on planet gap formation, 
focusing on parametrising the magnitude of gap opening with the gas/dust density contrast between the gap centre and gap edge. 
They concluded that it is generally harder for MHD winds to close up a planetary gap than viscosity for the same accretion rate, 
requiring $\alpha_{\rm dw}$ to be much larger than $\alpha_{\rm v}$ 
to maintain the same level of density depletion in the planetary gap. 
This implies for moderate values of $\alpha_{\rm dw}$, 
we can expect co-existence of low-$\alpha_{\rm v}$ controlled dust observational structure with high-$\alpha_{\rm dw}$ controlled accretion rate. 
When $\alpha_{\rm dw}\gg \alpha_{\rm v}$, 
disc wind may dominate both gap opening and accretion rates.

The majority of simulations from \citet{MHD-wind-Elbakyan} are in 1D and serve the purpose of parameter space exploration, 
supplemented by 2D two-fluid simulations for cross-validation, 
which contains gas and one dust species with a fixed Stokes number. 
In this work, 
we perform more realistic multi-fluid simulations with a physical dust size distribution, 
and post-process hydrodynamic simulation results to produce synthetic observations at ALMA wavelength and resolution. 
We focus on the full 2D shapes of dust brightness maps instead of only a hydrodynamic density contrast, 
and identify several characteristic features which demonstrate
the intrinsic differences in two kinds of accretion mechanisms. 

The paper is organized as follows: In \S \ref{sec: setup}, 
we introduce the numerical setup for performing hydrodynamic simulations and constructing synthetic images. 
In \S \ref{sec: results} we describe the disc morphology in characteristic regimes of the $\alpha_{\rm v}-\alpha_{\rm dw}$ parameter space. We summarise our findings and discuss the implications of our results in \S \ref{sec: discussion}.

\section{Numerical Setup}\label{sec: setup}

\subsection{Equations Solved}
\label{sec: wind-model}

In this work, we consider a 2D ($R$, $\varphi$), 
locally isothermal protoplanetary disc composed of gas and dust with an embedded planet. 
Disc self-gravity, 
realistic magnetic fields, 
planet orbital migration and accretion are not included in our simulations. 
The disc model is described by the Navier-Stokes equations in the cylindrical polar coordinate, 
in a corotating frame of reference centred on the central star:

\begin{equation}
    \frac{\partial\Sigma_{\rm g}}{\partial t}+\nabla\cdot\left(\Sigma_{\rm g}\textbf{\emph{V}}\right)=0\;,
    \label{HD-dens-gas}
\end{equation}
\begin{equation}
    \frac{\partial\textbf{\emph{V}}}{\partial t}+\left(\textbf{\emph{V}}\cdot\nabla\right)\textbf{\emph{V}}=-\nabla\Phi-\frac{\nabla P}{\Sigma_{\rm g}}-\dfrac{\sum\limits_{i}^{} \Sigma_{\rm d, \it i}\boldsymbol{F}_i}{\Sigma_{\rm g}}-\nabla\cdot\tau + \boldsymbol{S} \;,
    \label{HD-vel-gas}
\end{equation}

\begin{equation}
    \frac{\partial\Sigma_{\rm d, \it i}}{\partial t}+\nabla\cdot\left(\Sigma_{\rm g}\boldsymbol{W}_i+\boldsymbol{j_i}\right)=0\;,
    \label{HD-dens-dust}
\end{equation}
\begin{equation}
    \frac{\partial\boldsymbol{W}_i}{\partial t}+\left(\boldsymbol{W}_i\cdot\nabla\right)\boldsymbol{W}_i=-\nabla\Phi+\boldsymbol{F}_i\;,
    \label{HD-vel-dust}
\end{equation}

where $\Phi$ is the gravitational potential from the star and the planet, $\Sigma_{\rm g}$, 
$\textbf{\emph{V}}$ and $P$ are the surface density, 
velocity and pressure of the gas, 
respectively. 
$\Sigma_{\rm d, \it i}$ is the surface density for the $i$-th dust species, $\boldsymbol{W}_i$ is dust velocity, 
$\boldsymbol{F}_i$ is the drag force from gas acting on the dust of unit mass. $\boldsymbol{S}$ is the source term from wind accretion (parametrisation see below).

The viscous stress tensor $\tau$ is given by
\begin{equation}
\tau\equiv\Sigma_{\rm g}\nu\left[\nabla\boldsymbol{V}+\left(\nabla\boldsymbol{V}\right)^{T}-\frac{2}{3}\left(\nabla\cdot\boldsymbol{V}\right)\mathscr{I}\right],
\end{equation}
where $\mathscr{I}$ is the identity matrix, 
and 
the kinematic viscosity $\nu$ is specifically expressed as a function of a dimensionless constant  $\alpha_{\rm v}$ \citep{Shakura-Sunyaev73}:
\begin{equation}
    \nu=\alpha_{\rm v}c_{s}H,
\end{equation}

where $c_{s}$ is the sound speed, $H: = hR$ is the scale height, with $R$ being the radial distance and $h$ is the disc aspect ratio. 

Dust is treated as multiple pressure-less fluids. 
The drag force per mass for each species, $\boldsymbol{F}_i$, is given by
\begin{equation}
\boldsymbol{F}_i =\frac{\Omega_{\rm K}}{\St_i}\left(\boldsymbol{W}_i-\textbf{\emph{V}}\right),
\end{equation}

where $\Omega_{\rm K}$ is the Keplerian frequency, and the Stokes number (Epstein regime) for each species \citep{Birnstiel2010} proportional to the dust size is given by

\begin{equation}
\St_i =\frac{\pi}{2}\frac{s_{i}\rho_{\rm d}}{\Sigma_{\rm g}},
\end{equation}
where $s_i$ is grains size for different dust and $\rho_{\rm d}$ is the internal dust density.

For each species of dust $i$, our simulations also include dust diffusion flux $\boldsymbol{j}_i$:

\begin{equation}
\boldsymbol{j}_i = -D_{d,i}\left(\Sigma_{\rm g}+\Sigma_{\rm d, \it i}\right)\nabla\left(\frac{\Sigma_{\rm d, \it i}}{\Sigma_{\rm g}+\Sigma_{\rm d, \it i}}\right).
\end{equation}

Where $D_{d,i}$ is the dust diffusion coefficient \citep{Youdin-Lithwick}:

\begin{equation}
    D_{d,i} = \nu \dfrac{1+4\St_i^2}{(1+ \St_i^2)^2},
    \label{eqn:diffusion_coef}
\end{equation}

see \citet{Weber2019-dust-diffusion} for details on numerical implementation. 

To model the interplay between MHD wind and viscosity in our 2D hydrodynamic simulation, 
we modified the public code FARGO3D \citep{FARGO3D} in its multi-fluid version \citep{FARGO3D-multifluid}, 
see \citet{MHD-wind-Elbakyan}.
To summarise, 
we regard MHD disc winds 
as a specific torque $\Gamma$ acting to decrease 
the gas rotational angular momentum:

\begin{equation}
    \Gamma = \frac{1}{2} \sqrt{ \frac{GM_*}{R}} V_{\rm dw}\;,
    \label{Gamma}
\end{equation}

where $M_* = 1~M_{\odot}$ is the unit mass of the host star, 
and $V_{\rm dw}$ is the characteristic radial infall velocity of gas due to disc winds corresponding to this torque \citep{Tabone22}:

\begin{equation}
        V_{\rm dw} = -\frac{3}{2}\alpha_{\rm dw} h^2 V_{\rm K}\;,
    \label{v_dw}
\end{equation}

$\alpha_{\rm dw}$ is a dimensionless parameter related to the strength of the wind, 
defined similarly as the $\alpha_{\rm v}$ turbulent viscosity parameter, 
and $V_{\rm K}$ is the Keplerian velocity at $R$. 
It follows that the source term in Equation \ref{HD-vel-gas} would be

\begin{equation}
    \boldsymbol{S} = \Gamma \dfrac{\hat{\boldsymbol{R}}}{R},
\end{equation}

where ${\hat{\boldsymbol{R}}}$ is the unit vector for radial direction. 
Previous works used similar torque formulae to mimic wind-driven disc accretion in 2D hydrodynamic simulation, for example, \citet{Kimmig-2020} defined the inward velocity with a parameter $b$, 
roughly equivalent to $\pi \alpha_{\rm dw} h^2$ in our setup.
There is no dust diffusion associated with the disc wind.
\subsection{Disc Parameters}\label{sec: parameters}

The computational domain of our simulation ranges from $0.3\,R_0$ to $5.0\,R_0$, where $R_0$ the code unit length and planet orbital radius.
Although the hydrodynamic simulation is scale-free, we set $R_0=$~20~au for the purpose of synthetic map calculation, 
similar to \citet{2018ZhangDSHARP}. 
The simulation domain is resolved by 512 cells in the radial direction and 656 cells in the azimuthal direction. 
We apply the wave-damping boundary conditions as in the 2D simulations of \citet{MHD-wind-Elbakyan}.

The initial gas surface density profile is given by

\begin{equation}
    \Sigma_{\rm g}(R) = \Sigma_0 ({R}/{R_0})^{-1}\;,
    \label{sigma-disc0}
\end{equation}

where $\Sigma_0 = 2 \times 10^{-4} M_*/R_0^2$ in code units or $4.4 \text{g/cm}^2$ in physical units. 
For our local isothermal temperature profile, the aspect ratio is given by 

\begin{equation}
h(R) = c_s/V_{\rm K} = H/R = h_0 (R/R_0)^{0.25},
\end{equation}

where we adopt $h_0=0.05$ at $R_0$, 
corresponding to $T = 30$ K at 20~au for molecular hydrogen gas in physical units. 

We put 5 dust fluids with size $s$ spaced logarithmically uniformly between 30 $\mu$m and 3 mm. 
We choose an initial size distribution as $s^{-3.5}$ \citep{MRN}, 
an initial total dust-to-gas mass ratio $\epsilon=1\%$ and the internal grain density $\rho_{\rm d}=1.3~g/cm^{3}$. 
This gives us the Stokes number $\St\sim 0.03$ of mm-size dust for characteristic background density $\Sigma_0$. 

For a direct comparison, 
we illustrate the mass flux of pebbles transported by different processes in Figure \ref{fig:cartoon}. 
In both MHD windy discs and viscous discs, 
the planet tends to push the gas away from its vicinity \citep{Lin_Papaloizou_1993}. 
The local pressure minimum at the gas gap, therefore, 
drives the dust to drift away from the planet by gas drag \citep[the second term of the equation below, ][]{NakagawaEtal1986},

\begin{equation}\label{eq:W_dr}
    W_\mathrm{r} = \frac{1}{1+\mathrm{St}^2} V_{\rm r}+ \frac{\mathrm{1}}{\mathrm{St}+\mathrm{St}^{-1}} \frac{c_s^2}{\Omega_K r}\frac{\partial \log P}{\partial \log r},
\end{equation}

opening a dust gap as well (the red arrows in Figure \ref{fig:cartoon}). 
In the viscous discs, the turbulent diffusion of dust, whose net mass flux is along with the gas density gradient, aims to refill the gap (the purple arrows in Figure \ref{fig:cartoon}). 
However, in the MHD windy disc, wind only removes angular momentum thus gas velocity induced is always inward. 
In other words, the wind torque effect (Equation \ref{Gamma}) is equivalent to an additional gas radial velocity (Equation \ref{v_dw}), being always towards the host star. 
This gas radial velocity induced by wind can be comparable to or even larger than the gas viscous velocity depending on the value of $\alpha_{\rm dw}$, 
which directly influences the dust radial drift velocity (the first term of Equation \ref{eq:W_dr}), \textit{blowing} the pebble drift across the gap. 

In this work, 
we perform extensive simulations to explore the $\alpha_{\rm dw}$ and $\alpha_{\rm v}$ parameter space, 
while fixing the planet mass ratio $q = 3\times 10^{-4}$ (Saturn mass corresponding to a solar-mass host star) to isolate the effects of viscosity and wind on dust signatures. 
We choose this mass because it's massive enough to produce ALMA-like dusty gap signatures over a wide range of accretion parameters, 
and meanwhile consistent with the non-detection of super-Jupiter planets in ALMA sources by direct imaging campaigns \citep{Jorquera2021,Huelamo2022,Follette2022}.
See Figure \ref{fig:image} for all our choices of accretion parameters. 
Certain reference simulations have only the wind or the viscosity parameter.
Our simulations were run for 1000 orbital periods at $R_0$ ( 
corresponding to $\sim 0.1$ Myr for physical units), 
which is comparable to the typical disc ages of the DSHARP sources \citep{2018ZhangDSHARP}. 
All of the synthetic maps shown in this work refer to the configuration derived at this time.

\subsection{Radiative transfer}\label{sec: radmc-3d}

To generate synthetic maps for the dust continuum emission at sub-millimetre wavelength observable with ALMA, 
hydrodynamic results from FARGO3D multi-fluid simulations are 
post-processed with the ray-tracing algorithm in public code RADMC-3D \citep{RADMC-3D}. 
To extrapolate the 2D dust density map into 3D, we assume that all dust densities follow a Gaussian distribution in the vertical direction 
\begin{equation}
  \rho_{\rm d}(R,z) = \Sigma(R)\exp(-z^2/2H_{\rm d}^2)) / \sqrt{2\pi}.
  \label{eq:rho_d}
\end{equation}

Usually in the post-processing of images, one assumes $H_d \approx \sqrt{D_{\rm g}/(\Omega {\rm St})}$ \citep{Dubrulle1995,Youdin-Lithwick}, where $D_{\rm g}$ is the vertical turbulent diffusion coefficient at the midplane. One can relate $D_{\rm g} \approx \nu = \alpha_{\rm v}c_s H_{\rm g}$ for isotropic turbulence \citep{Birnstiel2016,Baruteau21} \footnote{Note that in the FARGO3D treatment of radial diffusion, Eqn \ref{eqn:diffusion_coef} has also implicitly made this assumption since $D_{\rm g}$ is put in place of $\nu$ in the original expression of \citet{Youdin-Lithwick}.}. However, MHD simulation results show this assumption is subject to many uncertainties \citep{Zhu-2015,Bai-2016}. The existence of wind may also bring dust to high altitudes, but that is usually for micron dust with very small Stokes number that already have $H_d$ intrinsically reaching the wind base at $\sim H_g$ \citep{Armitage2013}, not for millimeter size dust. 
Since by testing we found that for face-on disc, different vertical diffusion coefficient do not affect the final observation results, for simplicity and to focus on relatively face-on data analysis, 
the dust scale height $H_{\rm d}$ is fixed to $H_{\rm d} = 0.1~H_{\rm gas}$ for dust components similar to \citet{Miranda2017}, 
which is equivalent to a diffusivity coefficient $D_{\rm g} = 10^{-4}H_g^2\Omega$ for pebbles with stokes number of $0.01$. 
Such a dust scale height matches the ALMA observation in several discs \citep{Pinte2016,Villenave2020}, 
possibly implying different diffusivities of dust in the vertical and radial directions, {bf as found} in simulation with ideal or non-ideal MHD effects \citep{Bai2014,BaehrZhu2021,Zhou2022}. 

To compute realistic temperature structure of the disc based on dust density, 
we perform iterative thermal Monte Carlo simulations. 
Details of our procedure are essentially the same as described in \citet{Baruteau21}, 
although we apply opacities assuming compact dust species comprised of 70\% water ices and 30\% astro-silicates, 
with internal density $\rho_{\rm d}=1.3~g/cm^{3}$ consistent with our simulations, 
different from the porous dust ($\rho_{\rm d}=0.1~g/cm^{3}$) as assumed in \citet{fargo2radmc3d-dust}. 


To calculate the final beam-convolved synthetic images of dust continuum emission, we use the public python program \texttt{fargo2radmc3d} \citep{fargo2radmc3d-dust}. 
We assume the disc distance is 100 parsecs, and the stellar radius is 1.2 solar radii with a star effective temperature of 6000~K. 

We then compute the location-dependent emission flux of the disc at the wavelength of 1.25 mm (ALMA band~6). 
The brightness map roughly reflects the density distribution of dust with sizes close to this wavelength when the dust disc is generally optically thin. 
The final synthetic maps of the disc are then constructed by convolving the continuum intensity with a beam size of $0.02\times0.02$~arcsec (C\-9$\,+\,$C\-6 configuration pair) with an rms noise $10~\rm \mu Jy/beam$ (about $3$~hr on source time).

\section{Results}\label{sec: results}

We summarised all simulation parameters and results in Table \ref{table1}. 
For the naming of our models, for example, ``\texttt{v4dw1}'' means that there are both $\alpha_{\rm v} = 1\times10^{-4}$ and $\alpha_{\rm dw} = 1\times10^{-1}$ in this model.

\begin{figure*}
\centering
\includegraphics[width=0.99\linewidth]{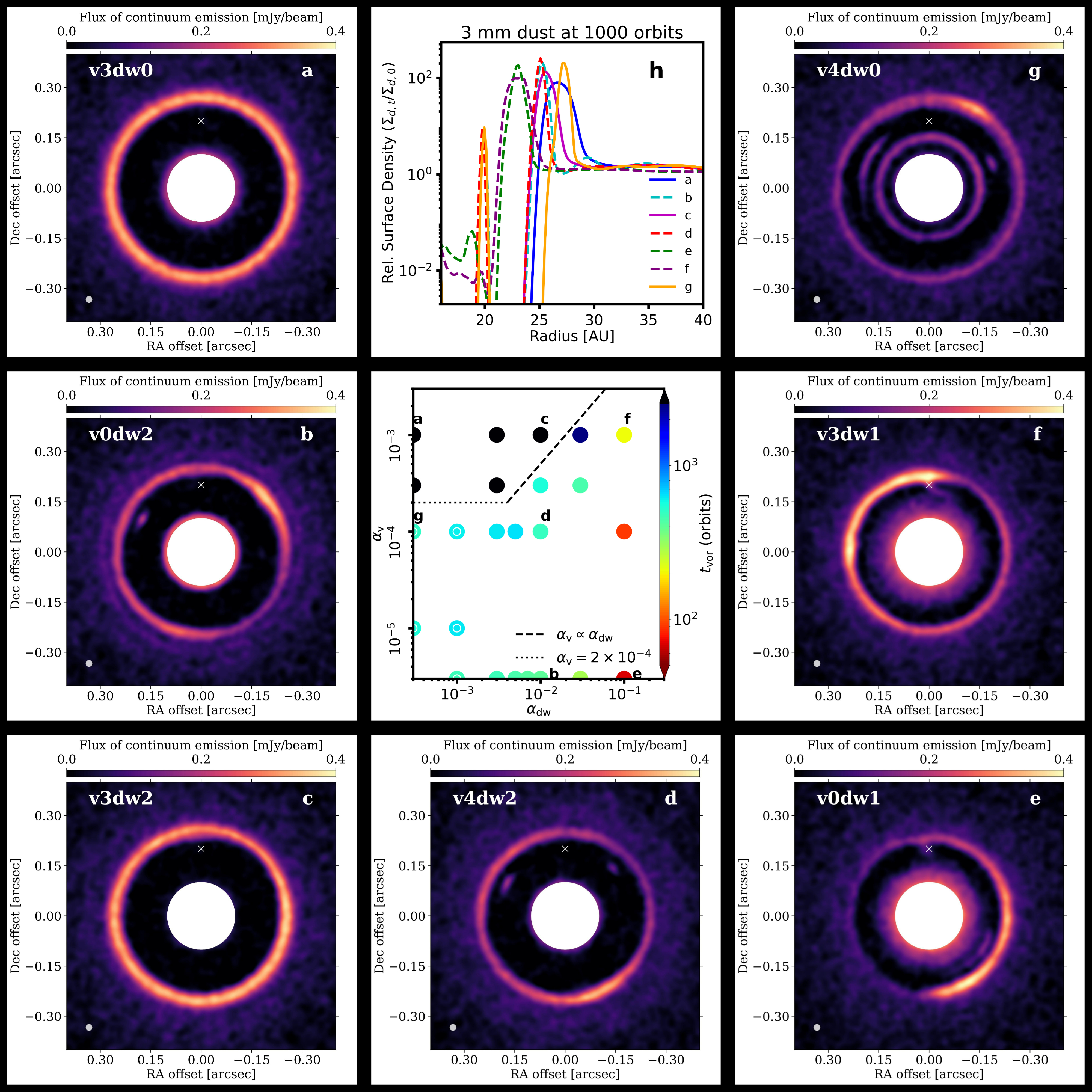}
\caption{\label{fig:image}The synthetic maps at the wavelength of 1.25mm, with the rms noise level $1.0 \times 10^{-5} \rm Jy/beam$. We select seven models with different $\alpha_{\rm v}$ and $\alpha_{\rm dw}$, obtained from hydrodynamic simulations with a planet (Saturn-mass) fixed at 20~au (the white “x” symbol represents the position of the planet). 
The model name and panel number are shown in the upper left and upper right corners of each panel, respectively. 
The beam size (20$\times$ 20\,mas) is shown in the lower-left corner of each image. The central plot shows the parameters ($\alpha_{\rm v}$ and $\alpha_{\rm dw}$), while $\alpha_{\rm v} =0.05 \alpha_{\rm dw}$ is plotted in dashed line. And we also plot the dotted line of  $\alpha_{\rm v} = 2\times10^{-4}$ to distinguish between high and low viscosity according to \citet{HammerEtal2021}. 
The choice of two accretion parameters in our simulation is plotted with colourful dots. Dots along the horizontal/vertical axes have one of the (subdominant) parameters set to zero. 
Different colours indicate different $t_{\rm vor}$ values as measured from simulations, it represents the moment of the first occurrence of vortex in each model. 
The white circle in the centre of the dot of some cases indicates that the case has an inner gap. The mid-upper panel (h) describes the radial profiles of 3mm dust species, 
normalized by their initial value at $R_0$, which the emission maps trace. 
Different line colours correspond to different models. 
These solid lines (viscosity) and dashed lines (wind) describe which physical process dominates the disc morphology formation in the vicinity of the planet.}
\end{figure*}

In the centre of Figure \ref{fig:image} we plot all our simulations as points in the $\alpha_{\rm v} - \alpha_{\rm dw}$ parameter space. 
Simulations with only the wind or viscosity parameter being non-zero are plotted on the horizontal/vertical axes.
The panels \texttt{a}-\texttt{g} of Figure \ref{fig:image} show selected synthetic emission maps for characteristic runs residing in different regions of the parameter space. 
As we will elaborate below, 
we find that asymmetry from the persistence of vortices is a significant feature that could differentiate a 
viscosity-dominated disc from a wind-dominated disc when the gap-ring profile in the emission map is quite similar \footnote{Here ``dominated" refers their relative control over characteristic features, 
not $\alpha_{\rm v} \gtrless \alpha_{\rm dw}$ in terms of accretion rate}. 
We use different colours in the middle panel in Figure \ref{fig:image} to represent the time vortices appear in our simulations. 
All symbols except black ones represent a run with vortices persisting to the end of the simulation.


\begin{table*}\textbf{{Hydrodynamical simulations models: parameters and results}}
\centering
\begin{tabular}{|c|c|c|c|c|c|c}\hline
Model&$\alpha_{\rm v}$&$\alpha_{\rm dw}$&$t_{\rm vor}$ (orbits)&Inner gap (Yes/No)&Synthetic maps (in Figure \ref{fig:image})\\
(1)&(2)&(3)&(4)&(5)&(6)
\\\hline
\multicolumn{6}{|c|}{\textbf{Strong-viscosity dominated}}\\
\hline\hline
\texttt{v3dw0}&$1\times10^{-3}$&0&$\infty$&No&a\\\hline
\texttt{v3dw(3.e-3)}&$1\times10^{-3}$&$3\times10^{-3}$&$\infty$&No&\\\hline
\multicolumn{6}{|c|}{\textbf{Moderate-viscosity dominated}}\\\hline\hline
\texttt{v3dw2}&$1\times10^{-3}$&$1\times10^{-2}$&$\infty$&No&c\\\hline
\texttt{v(3.e-4)dw(3.e-3)}&$3\times10^{-4}$&$3\times10^{-3}$&$\infty$&No&\\\hline
\multicolumn{6}{|c|}{\textbf{Moderate-wind dominated}}\\\hline\hline
\texttt{v0dw2}&0&$1\times10^{-2}$&410&No&b\\\hline
\texttt{v0dw(3.e-2)}&0&$3\times10^{-2}$&290&No&\\\hline
\texttt{v4dw2}&$1\times10^{-4}$&$1\times10^{-2}$&490&No&d\\\hline
\texttt{v(3.e-4)dw2}&$3\times10^{-4}$&$1\times10^{-2}$&550&No&\\\hline
\texttt{v(3.e-4)dw(3.e-2)}&$3\times10^{-4}$&$3\times10^{-2}$&440&No&\\\hline
\texttt{v3dw(3.e-2)}&$1\times10^{-3}$&$3\times10^{-2}$&$\approx 2500$&No&\\\hline
\texttt{v0dw1}&0&$1\times10^{-1}$&70&No&e\\\hline
\texttt{v4dw1}&$1\times10^{-4}$&$1\times10^{-1}$&90&No&\\\hline
\texttt{v3dw1}&$1\times10^{-3}$&$1\times10^{-1}$&210&No&f\\\hline
\texttt{v0dw(3.e-3)}&0&$3\times10^{-3}$&470&No&\\\hline
\texttt{v0dw(5.e-3)}&0&$5\times10^{-3}$&440&No&\\\hline
\texttt{v0dw(7.e-3)}&0&$7\times10^{-3}$&420&No&\\\hline
\texttt{v4dw(3.e-3)}&$1\times10^{-4}$&$3\times10^{-3}$&620&No&\\\hline
\texttt{v4dw(5.e-3)}&$1\times10^{-4}$&$5\times10^{-3}$&650&No&\\\hline
\multicolumn{6}{|c|}{\textbf{Inviscid disc}}\\
\hline\hline
\texttt{v5dw0}&$1\times10^{-5}$&0&550&Yes&\\\hline
\texttt{v5dw3}&$1\times10^{-5}$&$1\times10^{-3}$&620&Yes&\\\hline
\texttt{v4dw0}&$1\times10^{-4}$&0&500&Yes&g\\\hline
\texttt{v4dw3}&$1\times10^{-4}$&$1\times10^{-3}$&600&Yes&\\\hline
\texttt{v0dw3}&0&$1\times10^{-3}$&470&Yes&\\\hline
\end{tabular}
\caption{\label{table1}List of models adopted in our hydrodynamical simulations and shown in the middle panel in Figure \ref{fig:image}. We group the models which process dominate the emission feature in the simulations depending on the presence of asymmetries (the effects become inseparable in the inviscid regime), to narrow the range of the search upfront. Column (1) lists the name of the models.
The main two parameters $\alpha_{\rm v}$ and $\alpha_{\rm dw}$ are listed in Columns (2) and (3). Column (4) lists the simulation time of the vortex appearance $t_{\rm vor}$. 
Column (5) summarises whether there is a secondary gap in the corresponding model's synthetic map. Column (6) is a supplement to the understanding of Figure \ref{fig:image}.}
\end{table*}

\subsection{Moderate viscosity v.s. moderate wind accretion}\label{sec:3.1}

We first draw attention to one of our benchmark simulations with moderate viscosity $\alpha_{\rm v} = 10^{-3}$ and no wind \texttt{v3dw0} (Figure \ref{fig:image}a and blue line in Figure \ref{fig:image}h), 
in which we observe strongly symmetric gap+ring structures with no vortices. 
The bright dust ring is extended and its transition to the planetary gap is very smooth. 
Similar results are found in numerous previous works for discs with similar parameters \citep{Rosotti2016,2018ZhangDSHARP,facchini2018,2021Chen}. 

This is in contrast with wind-only simulations, all of which demonstrate visible asymmetries. 
In another benchmark simulation with moderate wind accretion parameter \texttt{v0dw2} (Figure \ref{fig:image}b), 
the planet is also able to open a deep gap in the dust profile, 
and with a gap width as \texttt{v3dw0} 
(Figure \ref{fig:image}a). However, the obvious clumps around ring for \texttt{v0dw2} (Figure \ref{fig:image}b) can help to easily distinguish the two models.

\begin{figure}
\centering
\includegraphics[width=0.49\hsize]{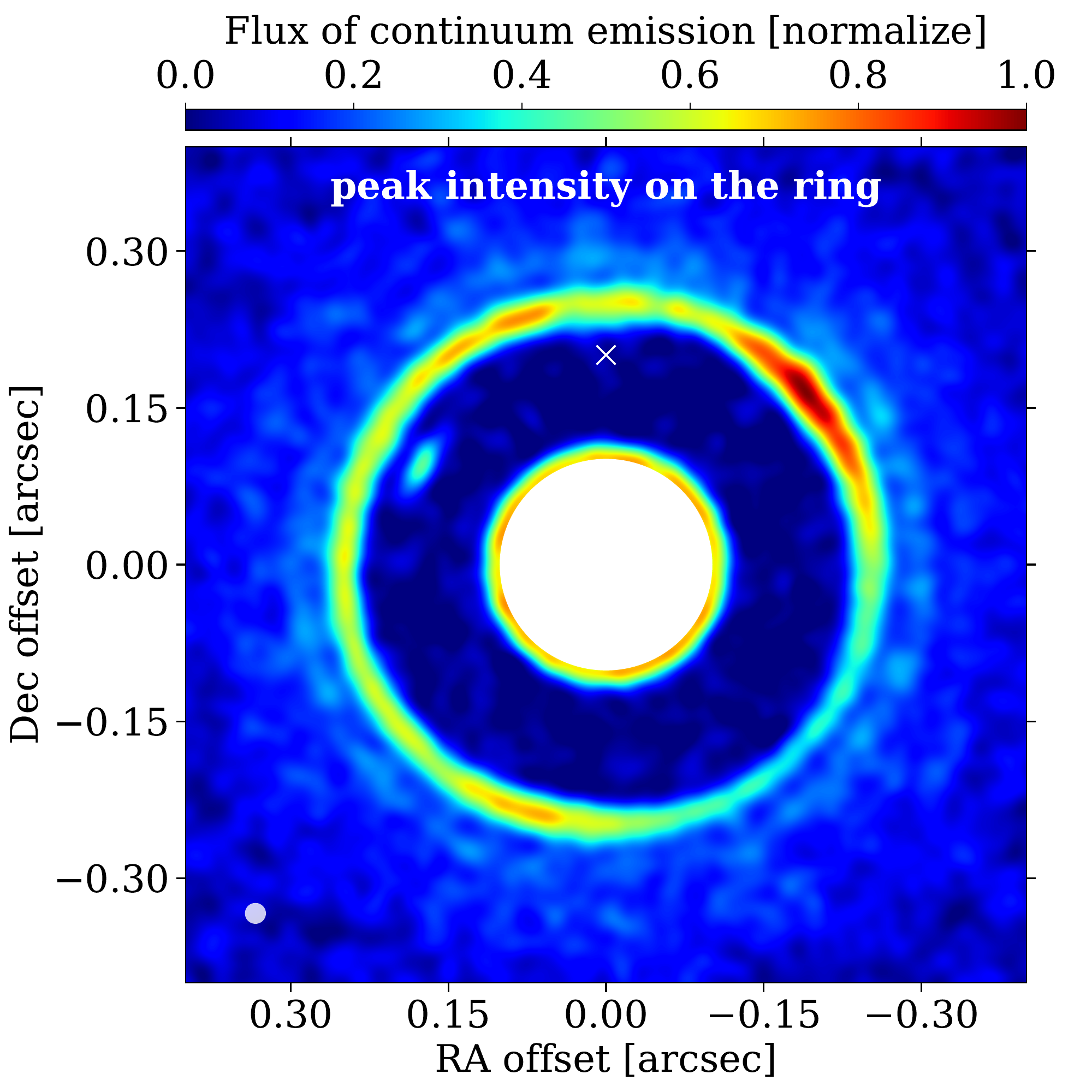}
\includegraphics[width=0.49\hsize]{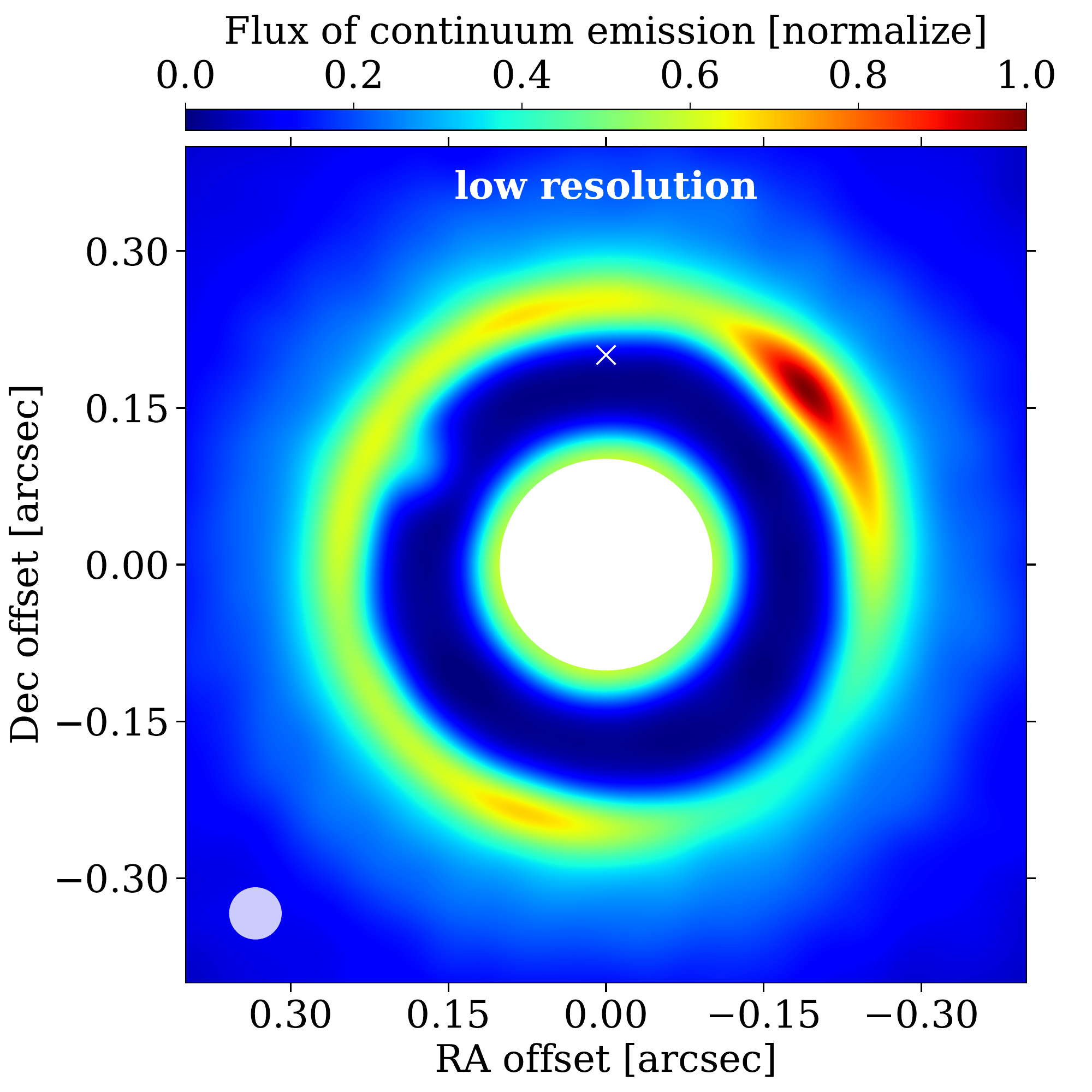}
\includegraphics[width=0.49\hsize]{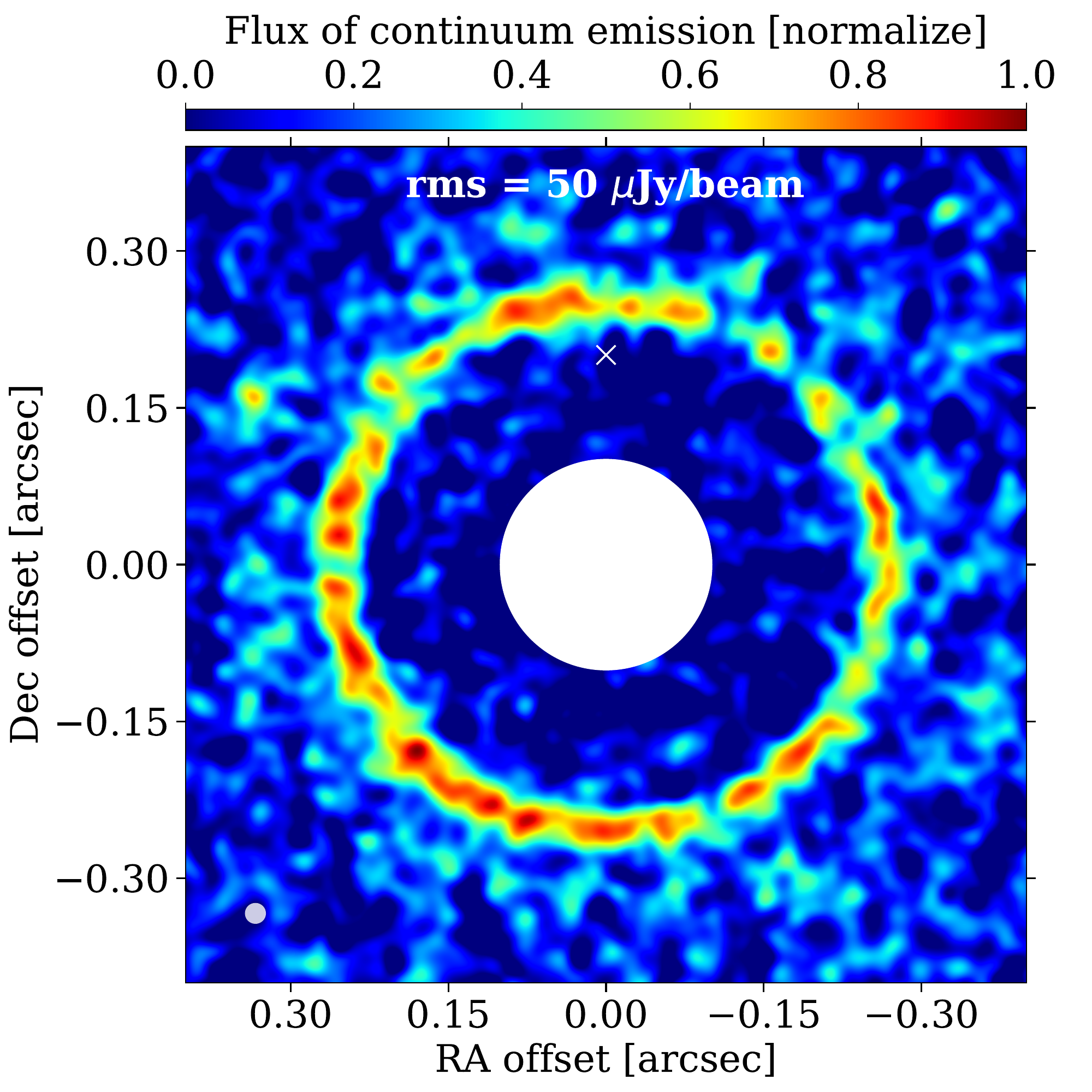}
\includegraphics[width=0.49\hsize]{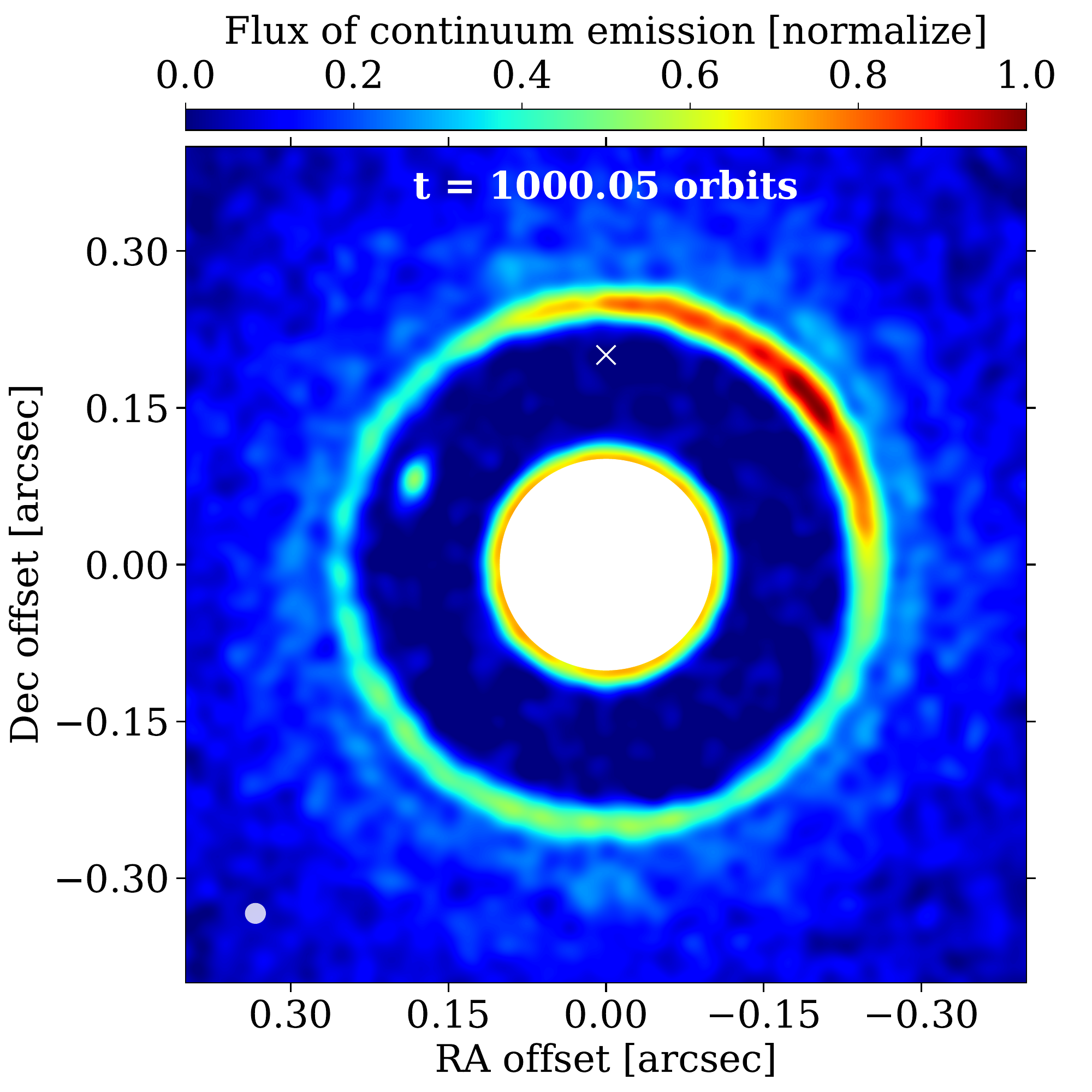}
\caption{\label{fig:normalize_image} Synthetic dust continuum maps for run \texttt{v0dw2}. The upper left panel takes the same setup (beam size 20$\times$20\,mas, rms noise $10~\rm \mu Jy/beam$) as Figure \ref{fig:image}b but with normalized intensity and linear colour bar. The spatial of the upper right panel is 50$\times$50\,mas. The rms noise for the lower left panel is $50~\rm \mu Jy/beam$. All setups of the lower right panel are the same as the upper left panel, but its snapshot time is 1000.05 orbits, approximately 5 years later.}
\end{figure}

To briefly investigate the influence of ALMA observation setups on our conclusions, 
we relax the spatial resolution and sensitivity requirement of the default setups in upper right and lower left panels in Figure \ref{fig:normalize_image}. 
Due to the azimuthal elongated shape (${\sim}90^\circ$), the ``detection" of the clump-like structures in \texttt{v0dw2} (Figure \ref{fig:image}b) is hardly affected when choosing the lower spatial resolution down to 50$\times$50\,mas (upper right panel of Figure \ref{fig:normalize_image}) as long as the ring feature can be resolved. 
However, the sensitivity of the simulated images plays a crucial role in the results. In the default setup, 
the signal-to-noise ratio of the clump peak is approximately 40. 
Increasing the rms noise from $10\rm \mu Jy/beam$ to $50~\rm \mu Jy/beam$ causes the asymmetric feature to become indistinguishable. 
This, in turn, suggests that clump-like features may be more commonly present on the pebble rings, 
as discussed in \S \ref{sec: discussion}.  Additionally, the clump-like feature is highly dynamic. 
In the lower right panel of Figure \ref{fig:normalize_image}, visible changes are apparent in the corotating frame when comparing the snapshot taken $\approx$5 years after the default image (approximately 0.05 orbits later).

\begin{figure}
\centering
\includegraphics[width=0.9\linewidth]{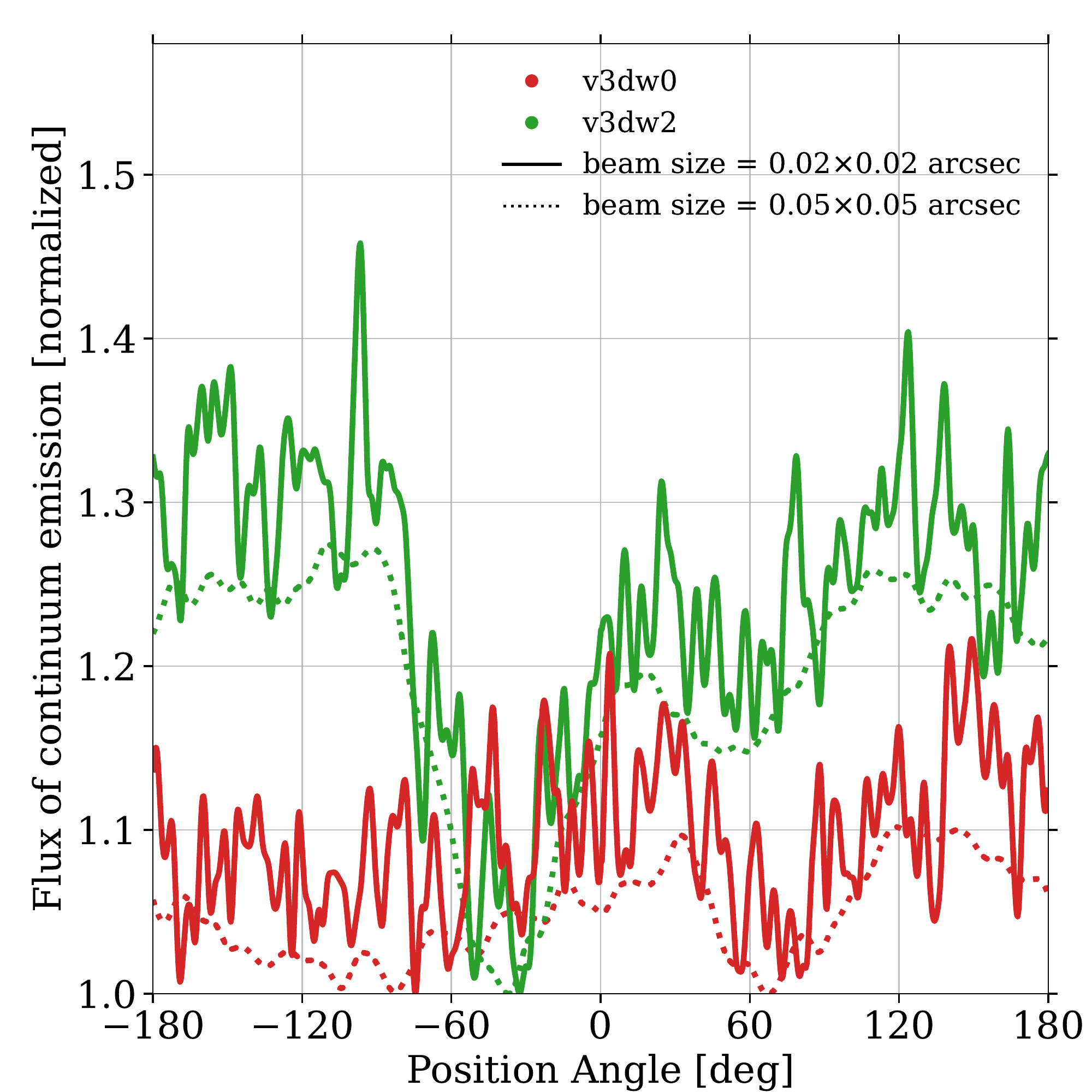}
\caption{\label{fig:flux}The radial average in their ring region (0.2$\sim$0.4 arcsec) of final intensity for \texttt{v3dw0} and \texttt{v3dw2}. The y-axis represents the normalized flux of continuum emission. Different colours correspond to different models. The solid lines represent the values obtained when the beam size is 20$\times$20\,mas. For comparison, we also plot the values obtained from low-resolution observations (beam size is 0.050 $\times$ 0.050 arcsec, dotted lines). The green-solid line is the upper left panel in Figure \ref{fig:normalize_image}, also shown in Figure \ref{fig:image}b. The green-dotted line is the upper right panel in Figure \ref{fig:normalize_image}. The red-solid line is also shown in Figure \ref{fig:image}a.}
\end{figure}

Where moderate wind and viscosity coexist, 
we find there is a transition between viscosity-dominated runs 
and wind-dominated runs. 
For example, 
when mild wind accretion 
$\alpha_{\rm dw} \leq 10^{-2}$ is added on top of \texttt{v3dw0} 
(run \texttt{v3dw2}, Figure \ref{fig:image}c), 
the disc is viscosity-dominated and the dust signature is still considerably symmetric.
If we now lower the viscosity to $\alpha_{\rm v} \leq 3\times 10^{-4}$ (e.g., run \texttt{v4dw2}, Figure \ref{fig:image}d), 
the brightness asymmetry would appear at around a few hundred orbits and persists for the rest of the simulation, 
similar to run \texttt{v0dw2}. Although it is hard to distinguish the \texttt{v3dw2} (Figure \ref{fig:image}c) from \texttt{v3dw0} (Figure \ref{fig:image}a) visually in the synthetic emission maps, 
their difference 
is still significant in terms of the azimuthal intensity variation.
In Figure \ref{fig:flux}, we normalize the intensity 
by the peak flux at the ring location. Whereas the ring in run \texttt{v3dw0} 
(Figure \ref{fig:image}a) shows fairly uniform distribution (red line) with $<10 \%$ fluctuation, 
up to 30\% variation is found in run \texttt{v3dw2} (Figure \ref{fig:image}b). 
Such a high contrast asymmetry on the ring could be easily tracked by ALMA observation. 

\begin{figure}
\centering
\includegraphics[width=0.49\hsize]{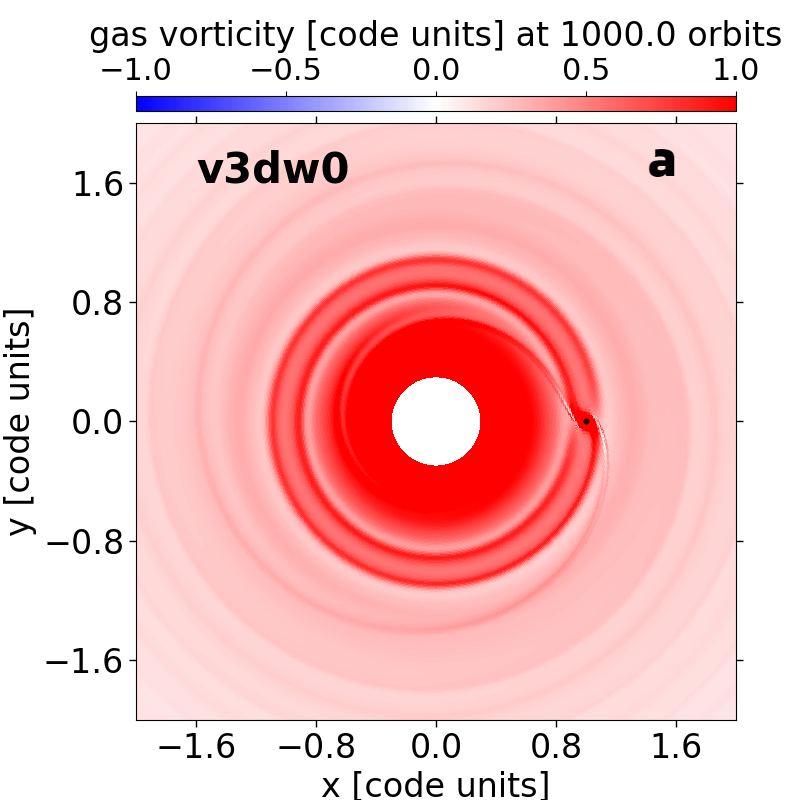}
\includegraphics[width=0.49\hsize]{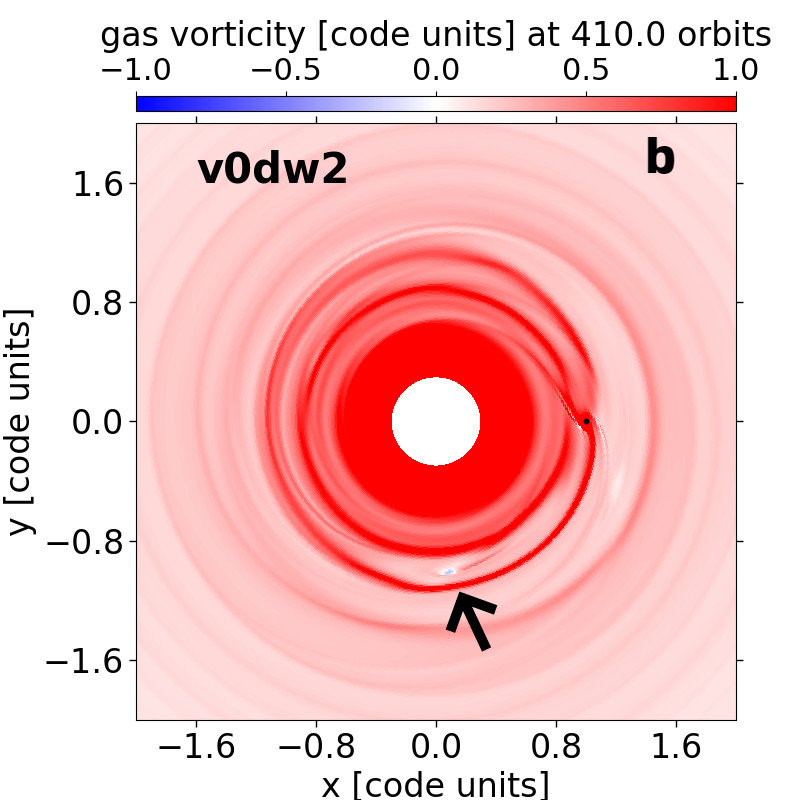}
\includegraphics[width=0.49\hsize]{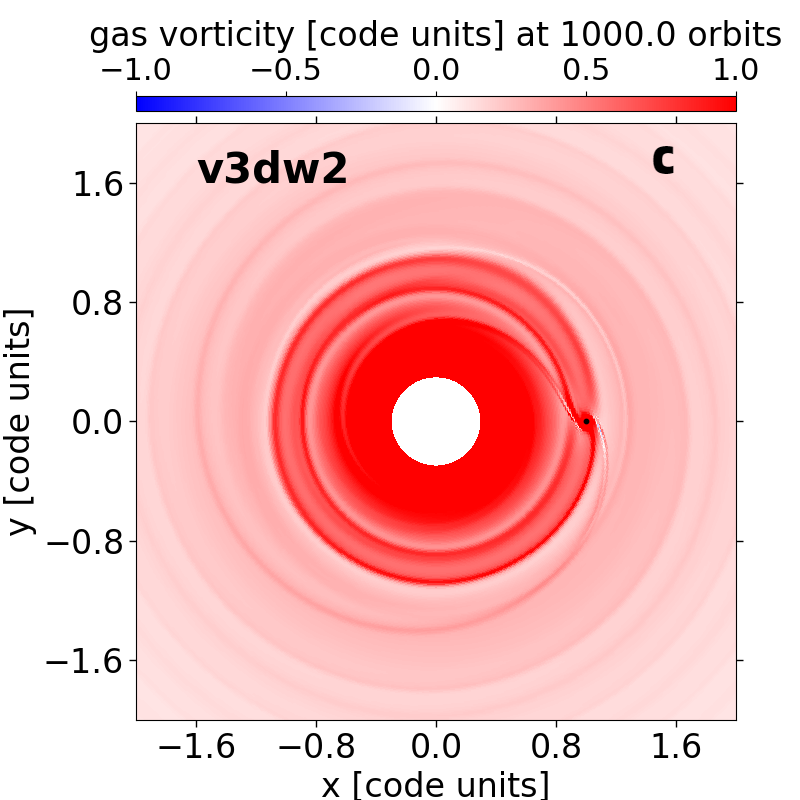}
\includegraphics[width=0.49\hsize]{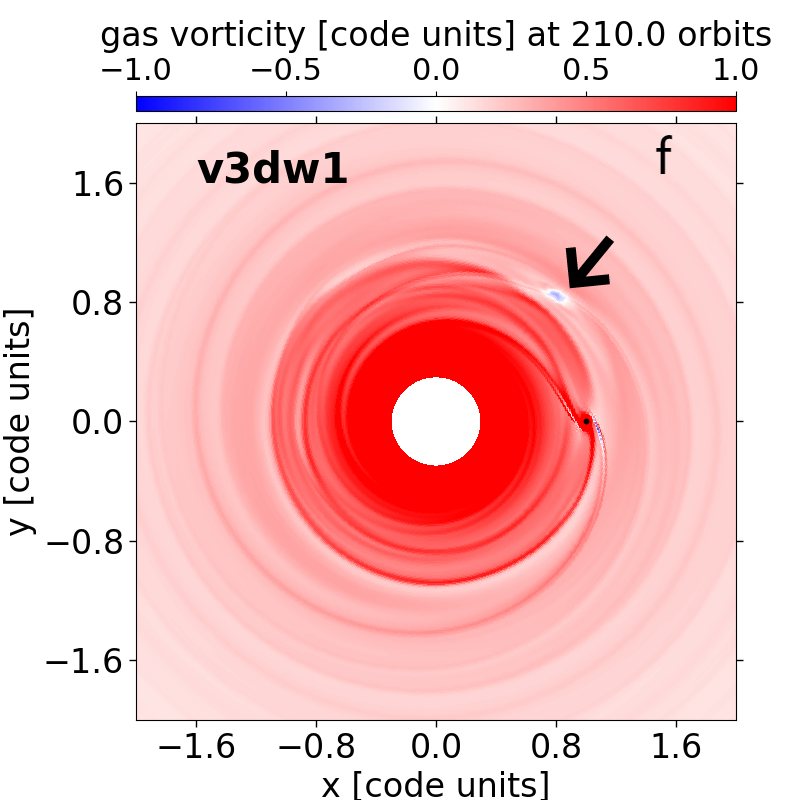}
\caption{\label{fig:vorticity}The snapshots of gas vorticity for different models (the panel number in Figure \ref{fig:image} corresponding to them is marked in the upper right corner). The top left corner of each panel displays the name of the model. The title of each panel contains the time of the corresponding snapshot. The black arrow points to the newly generated vortices.}
\end{figure}

This dichotomy is primarily associated with gas vortices on the gap outer edges being the source of asymmetries in wind-dominated runs. 
Vortices are generated from Rossby Wave Instability (RWI) \citep{RWI-1999,Li2001}, 
which results from steep gas surface density gradients at the edge of the planetary gaps.
Subsequently, 
they trap dust in clumps and lead to asymmetry in brightness maps.
To better quantify the production of the vortex, 
we define the appearing time of vortices in each model as $t_{\rm vor}$. 
The vortex is clearly visible in the 2D snapshots of the vorticity in our hydrodynamic simulation, 
in this way, 
the appearance of the vortex can be easily tracked. 
In Figure \ref{fig:vorticity}, 
we select snapshots of the vorticity of several models (\texttt{v3dw0}, \texttt{v0dw2}, \texttt{v3dw2}, and \texttt{v3dw1}), 
the black arrows point to the gas vortensity minimal points, 
i.e., the newly generated vortices. We plot $t_{\rm vor}$ for all models in the central panel of Figure
\ref{fig:image} in colours, which helps us to identify a transition from the wind-dominated parameter space towards the viscosity-dominated parameter space
(Complete simulation parameters and results for all models are presented in Table \ref{table1}).
Among them, 
black dots represent the models in which the vortex never disappears ($t_{\rm vor} = \infty$).
In general, 
$t_{\rm vor}$ increases from the bottom right to the upper left in the central panel of Figure \ref{fig:image}, 
which is closely related to the competition between $\alpha_{\rm v}$ and $\alpha_{\rm dw}$.
The steep change in $t_{\rm vor}$ in the moderate $\alpha_{\rm v}- \alpha_{\rm dw}$ parameter space 
sets up a natural boundary between wind-dominated (asymmetric) and viscosity-dominated regimes (symmetric).

\begin{figure}
\centering
\includegraphics[width=1.\linewidth]{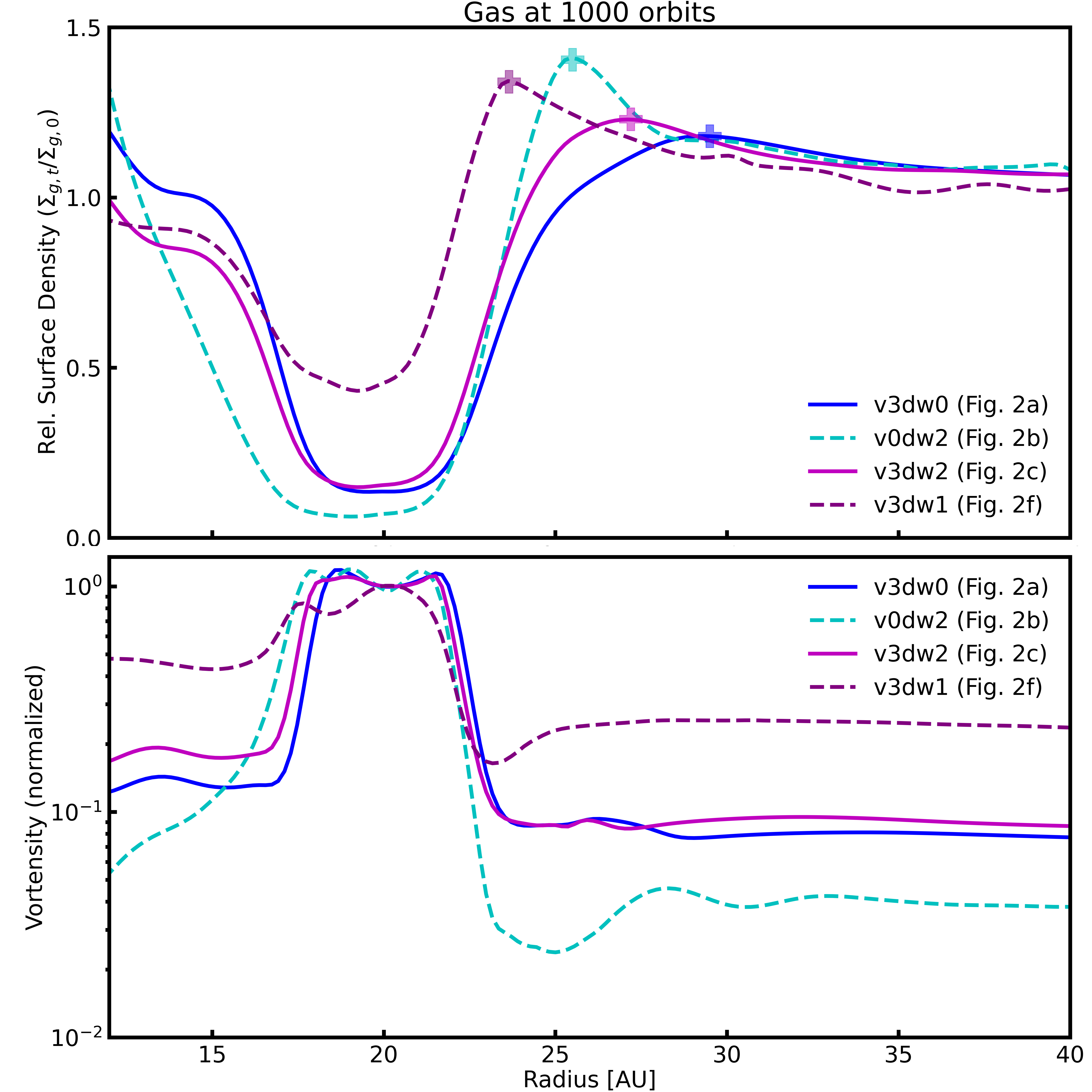}
\caption{\label{fig:dens_vortensity}Top row: radial profiles of the gas surface density normalized to the initial value. The crosses represent the peak value of the rings. Bottom row: radial profiles of the gas vortensity normalized to the value at $R_0$. The three lines in each panel correspond to three different models as labeled in each panel. As for how to distinguish between viscosity or wind-dominated model, we adopt the same expression as in Figure \ref{fig:image}h, that is, solid lines represent viscosity-dominated discs, and dotted lines represent wind-dominated discs. 
Both panels for disc models with a planet 20 au from the central star (\S \ref{sec: parameters}).}
\end{figure}

Why is disc wind constructive to vortex formation at the gap edges? 
Generally, 
steeper gas surface density gradients favour the onset of RWI. 
While both wind and viscosity 
tend to diminish the overall gas surface density contrast of a planetary gap \citep{MHD-wind-Elbakyan}, they seem to have competing effects on the maintenance of vortices. 
By looking into the radial gas distribution, we note that the existence of a local vortensity minima 
(condition for RWI) is more relevant to the smoothness of the density peak 
at the outer edge of the gap, 
rather than the overall density depletion at the planet's location.
For example, 
we plot the steady-state gas surface density profile (normalized by the initial unperturbed gas surface density) of \texttt{v3dw0}, \texttt{v0dw2}, \texttt{v3dw2}, and \texttt{v3dw1} (same as the models selected in Figure \ref{fig:vorticity}) in the upper panel of Figure \ref{fig:dens_vortensity}, 
with crosses indicating the closest density maxima outside the gap. 
The corresponding gas vortensity profiles are shown in the bottom panel.
The transition from the outer disc to the gap region in viscosity-only run \texttt{v3dw0} is very smooth, 
with a mild pressure maxima far away from the planet. 
This is because strong viscous diffusion smears out the vortensity gradient \citep{Li2001,Fu2014}.
The wind-only run \texttt{v0dw2} has a similar minimum density 
compared to \texttt{v3dw0} in the gap center, 
but the density bump outside the gap is much narrower and sharper, 
forming a conspicuous vortensity minimum.
This may be associated with disc winds providing an extra radial velocity to \textit{compress} the planet-induced pressure bump, 
forming a more peaked density profile at distances closer to the planet (thinner ring width and higher amplitude of the ring, both are preferred by the onset of RWI \citep{Ono2016}). 
Meanwhile, 
it does not contribute to the vortensity diffusion of gas or the concentration diffusion of dust. 
This is why even for the intermediate case \texttt{v3dw2}, 
while the gap centre is shallower than both \texttt{v3dw0} and \texttt{v0dw2}, 
the density bump is still more conspicuous compared to \texttt{v3dw0}, 
with a vortensity profile less monotonic (albeit the transition is not steep enough to generate RWI). 
Further increasing the $\alpha_{\rm dw}$ to $10^{-1}$ in \texttt{v3dw1},
will compress the gas ring even more and eventually lead to the onset of RWI and the prominent asymmetry as seen in panel f of Figure \ref{fig:image}, despite the gap center being filled.
Consistently, 
the location of density bumps in the dust profiles (Figure \ref{fig:image}h) show very similar trends. 
This demonstrates that wind and viscosity have opposite effects on the vortensity profile in the vicinity of the gap outer edge, 
leading to the symmetry dichotomy we observe in Figure \ref{fig:image}. 
It is intriguing to further scrutinize the time evolution of vortensity in these cases,
but since this paper is more focused on presenting results of a parameter survey 
and emphasising observable dichotomies, 
we will investigate the detailed mechanisms of wind compression 
in separate works.


To provide an order-of-magnitude description of wind-viscosity tension, 
suppose the pressure bump has a typical width of $\Delta = w R_0$, 
the timescale of viscous diffusion across the gap would be 

\begin{equation}\label{t_v}
    t_{\rm v} = \dfrac{\Delta^2}{\nu} = \dfrac{(w R_0)^2}{\alpha_{\rm v} h^2 V_K R_0},
\end{equation}

while the timescale of wind transport has a different dependence:

\begin{equation}\label{t_dw}
    t_{\rm dw} = \dfrac{\Delta}{V_{\rm dw}} = \dfrac{w R_0}{3\alpha_{\rm dw} h^2 V_K /2}.
\end{equation}

For viscous diffusion to be as effective as wind transport, 
we have $\alpha_{\rm v} \approx 3 w \alpha_{\rm dw}/2$. 
If $w$ is proportional to the dust gap width, usually determined by the planet Hill radius \citep{Rosotti2016,2021Chen}, 
then
$w \sim (q/3)^{1/3}$; 
Alternatively, it could be determined by the pressure scale height $w \sim h$, both of which do not depend on viscosity. 
In that case, 
we crudely expect the wind compression to out-balance the smoothing of vortensity by viscosity at $\alpha_{\rm v} \propto \alpha_{\rm dw}$, 
when $t_{\rm dw} < t_{\rm v}$. 

To conclude, wind-dominated discs with $w \alpha_{\rm dw} \geq \alpha_{v}$ show detectable asymmetry in their dust signatures due to the onset of RWI, 
which can be distinguished from viscosity-dominated symmetric discs with $w \alpha_{\rm dw} \leq \alpha_{v}$, where the viscosity is large enough $\alpha_{\rm v} \geq 2\times 10^{-4}$ to damp vortices by itself. 
In Figure \ref{fig:image} the separatrix on the wind viscosity plane between these regimes can be roughly fitted by $\alpha_{\rm v} \sim 0.05 \alpha_{\rm dw}$, 
in line with analytical predictions, since in our setup $(q/3)^{1/3} \sim h = 0.05$.

\subsection{Strong wind regime}

When the wind is relatively strong, e.g., \texttt{v0dw1} (Figure \ref{fig:image}e), 
profound asymmetry can be observed at the edge of a shallow and narrow gap. 
Shallow gaps in the traditional viscosity-dominated regime usually indicate either a small planet mass or higher viscosity. 
In such cases, 
the disc tends to exhibit more azimuthal symmetric features, as reported by \cite{Rosotti2016}. 
We note that shallow \& non-axisymmetric gaps may also be opened by eccentric planets \citep{2021Chen}, 
yet, 
the gap is usually wider in such a case. 
Since disc wind tends to fill up the planetary gap but also \textit{enhance} asymmetry, 
strong wind leads to a distinct feature consisting of a shallow and narrow gap within a non-axisymmetric dust ring, 
and even a coexisting moderate viscosity $\alpha_{\rm v} \sim 10^{-3}$ cannot damp the asymmetry features in run \texttt{v3dw1} 
(Figure \ref{fig:image}f). 
It is worth noting that a higher viscosity parameter of $\alpha_{\rm v}=10^{-2}$ could potentially dominate over wind and damp the asymmetry, 
resulting in a shallow and symmetric dust profile \citep{2018ZhangDSHARP}.
However,
such a level of viscosity is inconsistent with neither theoretical predictions nor ALMA observation constraints in protoplanetary discs \citep{2013BaiStone,Flaherty2017,Flaherty2020}.

A notable consequence of a stronger wind is its impact on the dust trapping at the ring, 
leading to leakier traps and ultimately resulting in the centre of the dust gap becoming shallower at higher $\alpha_{\rm dw}$. 
This effect is evident from the comparison of the radial profiles of \texttt{v3dw0}, \texttt{v3dw1}, and \texttt{v3dw2}, as well as \texttt{v4dw2} shown in Figure \ref{fig:image}h. 
The pebble leakage exacerbates the challenge of reconciling the dichotomy between carbonaceous (CC) and non-carbonaceous (NC) by giant planet gap opening \citep[e.g., ][]{StammlerEtal2023}. 
However, 
the leaked pebbles can potentially contribute to planet growth in the inner region, 
offering a possible explanation for the tentative correlation observed between super earth and exterior cold giant planet \citep{ZhuWu2018}.


Furthermore, 
the mass flux leaked by the strong wind can also alter the radial profile of the ring morphology 
(see Figure \ref{fig:image}h), 
leading to a ring steeper on the outer edge than the inner edge. 
This ring morphology is similar to the pebble ring held by traffic jams without significant gas substructures 
\citep{JiangOrmel2021}, 
but differs qualitatively from the planet-induced pebble ring in purely viscous discs, 
which has a steeper inner edge and flatter outer edge \citep[see][for a recent study]{BiEtal2023}, or see the profile of \texttt{v3dw0} in Figure \ref{fig:image}h for a comparison. 

Nevertheless, the typical disc lifetime for strong wind regime $\alpha_{\rm dw}\sim 0.1$ will be very short $\sim 10^4-10^5$ years \citep{Tabone22}, 
therefore we do not expect these features to be common in ALMA observations 
(see further discussion in \S \ref{sec: discussion}).

\subsection{Low viscosity v.s. Weak wind}

Our simulations focus on a range of viscosity parameters $\alpha_{\rm v}\sim 10^{-3}-10^{-4}$, 
which lie below the upper limit consistent with the dust diffusivity from theory and ALMA observation. 
For lower viscosity, $\alpha_{\rm v} \lesssim 10^{-4}$, the performance of the gas is nearly inviscid. Multiple rings and gaps can be present as demonstrated by \citet{Dong2017}. 
The detailed investigation on the inviscid regime ($\alpha_{\rm v} < 10^{-4}$ or $\alpha_{\rm dw} < 10^{-2}$) is not the main focus of this work, 
but the fact that low diffusivity prevents the destruction of the planet-induced vortex will help the understanding of our results.

For lower values of viscosity, 
the gaps are much deeper and the density contrasts much higher. 
Another important feature of gaps carved out by circular planets in low-viscosity environments is a secondary inner gap \citep{Dong2017,2018HuangDSHARP2}, 
appearing at $\sim H$. 
We verify that we have also observed this in our low-viscosity runs, 
see and the central panel of Figure \ref{fig:image} where we label an extra black circle on top of the symbol representing runs with inner rings in the lower-left parameter space (e.g., Figure \ref{fig:image}g, run \texttt{v4dw0}, as an example for synthetic maps). 
Since viscosity is low, 
the vortensity gradient cannot be smoothed even without MHD wind, 
and the vortex and asymmetry persist towards the end of simulations, 
as shown in \citet{HammerEtal2021,RometschEtal2021} for $\alpha_{\rm v} \leq 2\times 10^{-4}$. Eventually, 
we find that all cases in our simulations at low viscosity and weak wind regimes have inner gaps and asymmetries, 
this is consistent with the low viscosity simulations performed by previous authors \citep{Bae-2017}. 
In this context, 
the vortex cannot be a distinguishing factor between wind-dominated runs or viscosity-dominated runs, 
but these runs may still be separable from the rest of the parameter space by the inner ring feature. 
We note that the dust concentration at Lagrangian points L4 or L5 in certain Moderate-wind dominate cases or Inviscid cases seem to be a minor source for asymmetry, 
but since they are not maintained by gas vortices, 
they eventually disappear when we extend these runs to timescales longer by a few hundred orbits, as found in \citet{2018ZhangDSHARP,Long2022LkCa15disc}. Meanwhile, vortices in the outer rings are unaffected and remain to be the most essential source of asymmetry.

\section{Discussions \& Conclusions}\label{sec: discussion}

We propose a method to constrain the relative strength of the two angular momentum transfer mechanisms --- turbulent viscosity and magnetised disc wind --- in the planet-forming disc, by planet-induced dust continuum emission morphology. 
With a newly added parameterised MHD wind torque in FARGO3D \citep{Tabone22,MHD-wind-Elbakyan}, 
we perform a series of hydrodynamic simulations on planet gap opening with different levels of turbulent viscosity and disc wind, create mock ALMA images at 1.25mm, 
and test different sensitivity 
and spatial resolutions with radiation transfer code RADMC-3D. Our main conclusions are summarised as follows:

\begin{itemize}
    \item We divide the different observations that may be generated by the wind-viscosity interaction into four regimes (moderate-viscosity dominated, moderate-wind dominated, strong-wind dominated, inviscid), 
    characterised by the vortex appearance time and the disc morphology in the mock images of the ALMA continuum.

    \item We found that when strong wind dominates $\alpha_{\rm dw} \sim 0.1$, 
    the gap-opening planet produces a distinctive shallow gap and prominent asymmetric structure (vortex), 
    so we can directly distinguish it from the gap and ring morphology in the ALMA continuum. 
    Wind serves as an additional radial transport for both gas and dust, 
    making the gap shallower and altering the ring morphology.
    
    \item When moderate wind dominates $  \alpha_{\rm v}/w \leq \alpha_{\rm dw} \leq 0.01$ (where $w$ is the normalized width of the pressure bump at the outer edge of the planetary gap), 
    the high-contrast gap and ring feature it produces are similar to those of a traditional high-viscosity disc. 
    However, 
    there are obvious differences between the two cases in terms of the
    azimuthal intensity variation along the ring, 
    as the wind dominated case is considerably more asymmetric.
    
    \item In the two fore-mentioned regimes where wind transport is dominant, 
    vortices are easier to produce and can survive for a longer time, 
    even in the presence of viscosity damping. 
    We speculate this is because disc wind compression narrows the planet-generated gas surface density bump, 
    which results in a steeper vortensity transition 
    close to the outer edge of the gap and favours the generation of RWI. 

    \item When viscosity dominates $ \alpha_{\rm v} \geq w \alpha_{\rm dw}$, RWI can be damped and the disc will be highly symmetric, 
    separable from the wind dominated discs. 
    However, 
    for very low viscosity $\alpha_{\rm v} \leq 2\times 10^{-4}$, even viscous diffusion would become too weak to damp RWI, 
    and asymmetry will be ubiquitous in inviscid discs regardless of the wind strength. 
    Nevertheless, 
    such discs may be distinguishable from the rest of the parameter space with the presence of an inner secondary ring.
\end{itemize}

As the major novelty of this work, we highlight the enhancement of asymmetries by the wind. However, 
it is to be noted that our hydrodynamic simulations are limited by the 2D geometry, and MHD wind is simply added as a torque term that drives gas radial inflow. 
Such an approximation fails to capture the feedback effect of gap-opening process on the initial magnetic field structure or wind profile.
In a recent work by \citet{Aoyama-Bai-2023}, 
a detailed investigation of planet-disc interaction was conducted using 3D MHD simulations. They studied the process of MHD wind transporting gas across the planetary gap, 
and examined the migration torque that the disc exerts on the embedded planet. It is concluded that due to the concentration of magnetic flux in the planet vicinity, 
the wind torque within the gap is enhanced by a factor of 3-5. 
While these 3D MHD simulations are sophisticated and capture additional physics, they are computationally expensive and thus cannot be run for a long timescale that is relevant to ALMA disc ages. 
Furthermore, 
dust is not incorporated into the simulations to examine sub-mm emission. 
In comparison, 
our simulations are still capable of capturing the fundamental effects of viscosity and wind on the radial flow structure to first-order: while viscosity acts to close the gap on both sides, 
the wind compresses the gap from outer direction. 
Resorting to torque prescription and 2D geometry, 
we have the advantage of being able to run for much longer and across a broader range of parameters. 
We note that \citet[][see Appendix C]{Aoyama-Bai-2023} also conclude that Rossby wave instability persists for longer duration than typical viscous discs, 
albeit they did not explore whether pressure bump compression could occur when wind is introduced into a viscous simulation.
In future studies, 
it would be useful to incorporate more realistic prescriptions derived from sophisticated MHD simulations, 
such as an enhanced $\alpha_{\rm dw}$ in the gap center, into long-term and more efficient parameter surveys such as our own. 
Another caveat of our study is that we focused solely on face-on detection, 
which is insensitive to the vertical distribution of dust. 
In order to generate edge-on observations of planet-induced structures in wind-dominated discs, 
it may be necessary to conduct additional MHD disc wind simulations that incorporate dust, to provide a more self-consistent prescription of dust scale heights. 
However, 
disc wind launched above $\sim H_g$ may not significantly affect the vertical diffusion of mm-sized dust particles well-embedded in the midplane.

It is also shown by 2D \& 3D non-ideal MHD simulations that the large-scale poloidal field, required to launch MHD winds in the first place, may induce gap-like substructures in the absence of an embedded planet \citep{Riols2019,Suriano2019,Cui2021}. 
These substructures are also associated with pressure bumps that may act as dust traps. 
We anticipate that the compression of pressure bumps by wind is not limited to planet-induced pressure bumps, and therefore could lead to prolonged RWI and enhanced asymmetry in dust ring structures produced by other mechanisms as well.

Finally, we briefly discuss the asymmetry feature identified in our simulation in the context of archival ALMA observation. 
The very high-contrast asymmetry is hard to hide in ALMA observation. 
Prominent crescents \citep[e.g., IRS~48, MWC~758, HD135344B, HD~143006;][]{vanderMarelEtal2013a,MarinoEtal2015,vanderMarelEtal2016i,PerezEtal2018} may be relevant to the strong wind-dominated regime, 
albeit they are relatively rare \citep{vanderMarelEtal2021a}. 
This fact in turn suggests that $\alpha_{\rm dw}\sim 0.1$ might be not common in ALMA systems, 
which is also expected because their typical disc lifetime would be very short \citep{Tabone2022-2, Tabone22}.

However, in the moderate wind-dominant regime (\S \ref{sec:3.1}), 
the wind-driven asymmetry can still appear on top of an overall axisymmetric ring as clump-like features. 
Due to the relatively limited contrast between the clump peak and bottom, 
the very robust detection of these clump-like features might be challenging.
Meanwhile, 
depending on the geometry distribution of pebbles, 
the intensity along the major axis could be large than the minor axis of the ring simply due to the radiation transfer effect \citep{DoiKataoka2021}. 
More extremely, 
if the ring is optically thick, 
the ring could manifest an axisymmetry structure even with asymmetric azimuthal density profiles. 
We may expect to verify this in future facilities with longer wave band observation capability, e.g.,  the extended
ALMA array \citep[ALMA band 1,][]{Carpenter2020-future-ALMA, Burrill2022-future-ALMA}, the next-generation Very Large Array \citep[ngVLA,][]{Ricci2018-ngVLA} and the Square
Kilometre Array \citep[SKA,][]{Ilee2020-SKA, Wu2023-SKA}.

Yet, even though it is challenging, 
thanks to the improvement of spatial resolution of ALMA, 
a group of studies has pioneered identifying these clump-like structures on seemingly axisymmetric rings, 
e.g., the inner ring of the face-on disc HD~169142 \citep{MaciasEtal2017,PerezEtal2019}.
It has been proposed that the inner edge of a circumbinary disc can feature these clump structures \citep{PobleteEtal2022}. Alternatively, 
these clump features can also be caused by mesoscale instability triggered by dust feedback in massive rings \citep{HuangEtal2020p}. Contradictory results are obtained by \citet{LehmannLin2022}, who suggest that a higher-metallicity disc can kill the short-lived vortex more easily, 
and smear out the asymmetry. 
Our result suggests it could alternatively be achieved if disc wind dominates over viscosity in the inner disc of HD~169142. 

Another potential target of interest is RXJ1604.3–2130~A, where, again, 
the significant level ($\sim 30\%$) of intensity variation is observed in the continuum of the transitional disc rim \citep{StadlerEtal2023}. 
It is proposed that the variation could be caused by the shadowing of an inclined inner disc, 
as in the near-infrared counterpart of RXJ1604.3–2130~A \citep{PinillaEtal2018}, 
while this assumption largely depends on the uncertain disc cooling. 

In the end, the intensity of the continuum ring beyond the orbit of PDS~70~b and c exhibits some moderate asymmetry in the azimuthal directions \citep{KepplerEtal2019}, 
which might be potentially explained by vortices at the ring location. 
As there are two planets confirmed in PDS 70, 
it will be appealing to compare our results with the PDS 70 ALMA continuum, 
as well as explore detailed dependencies of ring morphology on planet mass, 
in subsequent works.

\section*{Acknowledgements}
We thank our referee, Can Cui, for a set of useful comments that improved the presentation of this work. We thank Xue-Ning Bai, Kaitlin Kratter, Sergei Nayakshin, Chris W. Ormel and Martin Pessah for their helpful comments and discussions. This research used the DiRAC Data Intensive service at Leicester, operated by the University of Leicester IT Services, which forms part of the STFC DiRAC HPC Facility (\href{www.dirac.ac.uk}{www.dirac.ac.uk}). 

Y.W. gratefully acknowledges Heising-Simons Foundation for funding his travel in ERES VIII and the support by the DUSTBUSTERS RISE project (grant agreement number 823823) for his secondment at the University of Arizona. M-K.L. acknowledges the funding from the National Science and Technology Council through grants 110-2112-M-001-034-, 111-2112-M-001-062-, 111-2124-M-002-013-, 112-2124-M-002-003- and an Academia Sinica Career Development Award (AS-CDA-110-M06). G.R. acknowledges support from an STFC Ernest Rutherford Fellowship (grant number ST/T003855/1) and is funded by the European Union (ERC DiscEvol, project number 101039651). Views and opinions expressed are however those of the author(s) only and do not necessarily reflect those of the European Union or the European Research Council Executive Agency. Neither the European Union nor the granting authority can be held responsible for them. V.E. acknowledges the funding from the UK Science and Technologies Facilities Council, grant No. ST/S000453/1 and the support of the Ministry of Science and Higher Education of the Russian Federation (State assignment in the field of scientific activity 2023, GZ0110/23-10-IF). 

\section*{Data availability}

The data obtained in our simulations can be made available on reasonable request to the corresponding author. 




\bibliographystyle{mnras}
\bibliography{multiplanet} 

\begin{thebibliography}{}
\makeatletter
\relax
\def\mn@urlcharsother{\let\do\@makeother \do\$\do\&\do\#\do\^\do\_\do\%\do\~}
\def\mn@doi{\begingroup\mn@urlcharsother \@ifnextchar [ {\mn@doi@}
  {\mn@doi@[]}}
\def\mn@doi@[#1]#2{\def\@tempa{#1}\ifx\@tempa\@empty \href
  {http://dx.doi.org/#2} {doi:#2}\else \href {http://dx.doi.org/#2} {#1}\fi
  \endgroup}
\def\mn@eprint#1#2{\mn@eprint@#1:#2::\@nil}
\def\mn@eprint@arXiv#1{\href {http://arxiv.org/abs/#1} {{\tt arXiv:#1}}}
\def\mn@eprint@dblp#1{\href {http://dblp.uni-trier.de/rec/bibtex/#1.xml}
  {dblp:#1}}
\def\mn@eprint@#1:#2:#3:#4\@nil{\def\@tempa {#1}\def\@tempb {#2}\def\@tempc
  {#3}\ifx \@tempc \@empty \let \@tempc \@tempb \let \@tempb \@tempa \fi \ifx
  \@tempb \@empty \def\@tempb {arXiv}\fi \@ifundefined
  {mn@eprint@\@tempb}{\@tempb:\@tempc}{\expandafter \expandafter \csname
  mn@eprint@\@tempb\endcsname \expandafter{\@tempc}}}

\bibitem[\protect\citeauthoryear{{ALMA Partnership} et~al.,}{{ALMA Partnership}
  et~al.}{2015}]{ALMA2015}
{ALMA Partnership} et~al., 2015, \mn@doi [\apjl] {10.1088/2041-8205/808/1/L3},
  \href {https://ui.adsabs.harvard.edu/abs/2015ApJ...808L...3A} {808, L3}

\bibitem[\protect\citeauthoryear{{Andrews}}{{Andrews}}{2020}]{Andrews2020}
{Andrews} S.~M.,  2020, \mn@doi [\araa] {10.1146/annurev-astro-031220-010302},
  \href {https://ui.adsabs.harvard.edu/abs/2020ARA&amp;A..58..483A} {58, 483}

\bibitem[\protect\citeauthoryear{{Andrews} et~al.,}{{Andrews}
  et~al.}{2018}]{andrews2018}
{Andrews} S.~M.,  et~al., 2018, \mn@doi [\apjl] {10.3847/2041-8213/aaf741},
  \href {https://ui.adsabs.harvard.edu/abs/2018ApJ...869L..41A} {869, L41}

\bibitem[\protect\citeauthoryear{{Aoyama} \& {Bai}}{{Aoyama} \&
  {Bai}}{2023}]{Aoyama-Bai-2023}
{Aoyama} Y.,  {Bai} X.,  2023, \mn@doi [arXiv e-prints]
  {10.48550/arXiv.2302.01514}, \href
  {https://ui.adsabs.harvard.edu/abs/2023arXiv230201514A} {p. arXiv:2302.01514}

\bibitem[\protect\citeauthoryear{{Armitage}, {Simon}  \& {Martin}}{{Armitage}
  et~al.}{2013}]{Armitage2013}
{Armitage} P.~J.,  {Simon} J.~B.,   {Martin} R.~G.,  2013, \mn@doi [\apjl]
  {10.1088/2041-8205/778/1/L14}, \href
  {https://ui.adsabs.harvard.edu/abs/2013ApJ...778L..14A} {778, L14}

\bibitem[\protect\citeauthoryear{{Bae}, {Zhu}  \& {Hartmann}}{{Bae}
  et~al.}{2017}]{Bae-2017}
{Bae} J.,  {Zhu} Z.,   {Hartmann} L.,  2017, \mn@doi [\apj]
  {10.3847/1538-4357/aa9705}, \href
  {https://ui.adsabs.harvard.edu/abs/2017ApJ...850..201B} {850, 201}

\bibitem[\protect\citeauthoryear{{Baehr} \& {Zhu}}{{Baehr} \&
  {Zhu}}{2021}]{BaehrZhu2021}
{Baehr} H.,  {Zhu} Z.,  2021, \mn@doi [\apj] {10.3847/1538-4357/abddb4}, \href
  {https://ui.adsabs.harvard.edu/abs/2021ApJ...909..136B} {909, 136}

\bibitem[\protect\citeauthoryear{{Bai} \& {Stone}}{{Bai} \&
  {Stone}}{2013}]{2013BaiStone}
{Bai} X.-N.,  {Stone} J.~M.,  2013, \mn@doi [\apj]
  {10.1088/0004-637X/769/1/76}, \href
  {https://ui.adsabs.harvard.edu/abs/2013ApJ...769...76B} {769, 76}

\bibitem[\protect\citeauthoryear{{Bai} \& {Stone}}{{Bai} \&
  {Stone}}{2014}]{Bai2014}
{Bai} X.-N.,  {Stone} J.~M.,  2014, \mn@doi [\apj]
  {10.1088/0004-637X/796/1/31}, \href
  {https://ui.adsabs.harvard.edu/abs/2014ApJ...796...31B} {796, 31}

\bibitem[\protect\citeauthoryear{{Bai}, {Ye}, {Goodman}  \& {Yuan}}{{Bai}
  et~al.}{2016}]{Bai-2016}
{Bai} X.-N.,  {Ye} J.,  {Goodman} J.,   {Yuan} F.,  2016, \mn@doi [\apj]
  {10.3847/0004-637X/818/2/152}, \href
  {https://ui.adsabs.harvard.edu/abs/2016ApJ...818..152B} {818, 152}

\bibitem[\protect\citeauthoryear{{Baruteau} et~al.,}{{Baruteau}
  et~al.}{2019}]{fargo2radmc3d-dust}
{Baruteau} C.,  et~al., 2019, \mn@doi [\mnras] {10.1093/mnras/stz802}, \href
  {https://ui.adsabs.harvard.edu/abs/2019MNRAS.486..304B} {486, 304}

\bibitem[\protect\citeauthoryear{{Baruteau}, {Wafflard-Fernandez}, {Le Gal},
  {Debras}, {Carmona}, {Fuente}  \& {Rivi{\`e}re-Marichalar}}{{Baruteau}
  et~al.}{2021}]{Baruteau21}
{Baruteau} C.,  {Wafflard-Fernandez} G.,  {Le Gal} R.,  {Debras} F.,  {Carmona}
  A.,  {Fuente} A.,   {Rivi{\`e}re-Marichalar} P.,  2021, \mn@doi [\mnras]
  {10.1093/mnras/stab1045}, \href
  {https://ui-adsabs-harvard-edu.insu.bib.cnrs.fr/abs/2021MNRAS.505..359B}
  {505, 359}

\bibitem[\protect\citeauthoryear{{Ben{\'\i}tez-Llambay} \&
  {Masset}}{{Ben{\'\i}tez-Llambay} \& {Masset}}{2016}]{FARGO3D}
{Ben{\'\i}tez-Llambay} P.,  {Masset} F.~S.,  2016, \mn@doi [\apjs]
  {10.3847/0067-0049/223/1/11}, \href
  {https://ui.adsabs.harvard.edu/abs/2016ApJS..223...11B} {223, 11}

\bibitem[\protect\citeauthoryear{{Ben{\'\i}tez-Llambay}, {Krapp}  \&
  {Pessah}}{{Ben{\'\i}tez-Llambay} et~al.}{2019}]{FARGO3D-multifluid}
{Ben{\'\i}tez-Llambay} P.,  {Krapp} L.,   {Pessah} M.~E.,  2019, \mn@doi
  [\apjs] {10.3847/1538-4365/ab0a0e}, \href
  {https://ui.adsabs.harvard.edu/abs/2019ApJS..241...25B} {241, 25}

\bibitem[\protect\citeauthoryear{{Bi}, {Lin}  \& {Dong}}{{Bi}
  et~al.}{2023}]{BiEtal2023}
{Bi} J.,  {Lin} M.-K.,   {Dong} R.,  2023, \mn@doi [\apj]
  {10.3847/1538-4357/aca1b1}, \href
  {https://ui.adsabs.harvard.edu/abs/2023ApJ...942...80B} {942, 80}

\bibitem[\protect\citeauthoryear{{Birnstiel}, {Dullemond}  \&
  {Brauer}}{{Birnstiel} et~al.}{2010}]{Birnstiel2010}
{Birnstiel} T.,  {Dullemond} C.~P.,   {Brauer} F.,  2010, \mn@doi [\aap]
  {10.1051/0004-6361/200913731}, \href
  {https://ui.adsabs.harvard.edu/abs/2010A&A...513A..79B} {513, A79}

\bibitem[\protect\citeauthoryear{{Birnstiel}, {Fang}  \&
  {Johansen}}{{Birnstiel} et~al.}{2016}]{Birnstiel2016}
{Birnstiel} T.,  {Fang} M.,   {Johansen} A.,  2016, \mn@doi [\ssr]
  {10.1007/s11214-016-0256-1}, \href
  {https://ui.adsabs.harvard.edu/abs/2016SSRv..205...41B} {205, 41}

\bibitem[\protect\citeauthoryear{{Burrill}, {Ricci}, {Harter}, {Zhang}  \&
  {Zhu}}{{Burrill} et~al.}{2022}]{Burrill2022-future-ALMA}
{Burrill} B.~P.,  {Ricci} L.,  {Harter} S.~K.,  {Zhang} S.,   {Zhu} Z.,  2022,
  \mn@doi [\apj] {10.3847/1538-4357/ac5592}, \href
  {https://ui.adsabs.harvard.edu/abs/2022ApJ...928...40B} {928, 40}

\bibitem[\protect\citeauthoryear{{Carpenter}, {Iono}, {Kemper}  \&
  {Wootten}}{{Carpenter} et~al.}{2020}]{Carpenter2020-future-ALMA}
{Carpenter} J.,  {Iono} D.,  {Kemper} F.,   {Wootten} A.,  2020, \mn@doi [arXiv
  e-prints] {10.48550/arXiv.2001.11076}, \href
  {https://ui.adsabs.harvard.edu/abs/2020arXiv200111076C} {p. arXiv:2001.11076}

\bibitem[\protect\citeauthoryear{{Chen}, {Wang}, {Li}, {Baruteau}  \&
  {Lin}}{{Chen} et~al.}{2021}]{2021Chen}
{Chen} Y.-X.,  {Wang} Z.,  {Li} Y.-P.,  {Baruteau} C.,   {Lin} D. N.~C.,  2021,
  \mn@doi [\apj] {10.3847/1538-4357/ac23d7}, \href
  {https://ui.adsabs.harvard.edu/abs/2021ApJ...922..184C} {922, 184}

\bibitem[\protect\citeauthoryear{{Cui} \& {Bai}}{{Cui} \&
  {Bai}}{2021}]{Cui2021}
{Cui} C.,  {Bai} X.-N.,  2021, \mn@doi [\mnras] {10.1093/mnras/stab2220}, \href
  {https://ui.adsabs.harvard.edu/abs/2021MNRAS.507.1106C} {507, 1106}

\bibitem[\protect\citeauthoryear{{Cui} \& {Bai}}{{Cui} \&
  {Bai}}{2022}]{Cui2022}
{Cui} C.,  {Bai} X.-N.,  2022, \mn@doi [\mnras] {10.1093/mnras/stac2580}, \href
  {https://ui.adsabs.harvard.edu/abs/2022MNRAS.516.4660C} {516, 4660}

\bibitem[\protect\citeauthoryear{{Doi} \& {Kataoka}}{{Doi} \&
  {Kataoka}}{2021}]{DoiKataoka2021}
{Doi} K.,  {Kataoka} A.,  2021, \mn@doi [\apj] {10.3847/1538-4357/abe5a6},
  \href {https://ui.adsabs.harvard.edu/abs/2021ApJ...912..164D} {912, 164}

\bibitem[\protect\citeauthoryear{{Dong}, {Li}, {Chiang}  \& {Li}}{{Dong}
  et~al.}{2017}]{Dong2017}
{Dong} R.,  {Li} S.,  {Chiang} E.,   {Li} H.,  2017, \mn@doi [\apj]
  {10.3847/1538-4357/aa72f2}, \href
  {https://ui.adsabs.harvard.edu/abs/2017ApJ...843..127D} {843, 127}

\bibitem[\protect\citeauthoryear{{Dubrulle}, {Morfill}  \&
  {Sterzik}}{{Dubrulle} et~al.}{1995}]{Dubrulle1995}
{Dubrulle} B.,  {Morfill} G.,   {Sterzik} M.,  1995, \mn@doi [\icarus]
  {10.1006/icar.1995.1058}, \href
  {https://ui.adsabs.harvard.edu/abs/1995Icar..114..237D} {114, 237}

\bibitem[\protect\citeauthoryear{{Dullemond}, {Juhasz}, {Pohl}, {Sereshti},
  {Shetty}, {Peters}, {Commercon}  \& {Flock}}{{Dullemond}
  et~al.}{2012}]{RADMC-3D}
{Dullemond} C.~P.,  {Juhasz} A.,  {Pohl} A.,  {Sereshti} F.,  {Shetty} R.,
  {Peters} T.,  {Commercon} B.,   {Flock} M.,  2012, {RADMC-3D: A multi-purpose
  radiative transfer tool} (\mn@eprint {ascl} {1202.015})

\bibitem[\protect\citeauthoryear{{Dullemond} et~al.,}{{Dullemond}
  et~al.}{2018}]{Dullemond2018}
{Dullemond} C.~P.,  et~al., 2018, \mn@doi [\apjl] {10.3847/2041-8213/aaf742},
  \href {https://ui.adsabs.harvard.edu/abs/2018ApJ...869L..46D} {869, L46}

\bibitem[\protect\citeauthoryear{{Elbakyan}, {Wu}, {Nayakshin}  \&
  {Rosotti}}{{Elbakyan} et~al.}{2022}]{MHD-wind-Elbakyan}
{Elbakyan} V.,  {Wu} Y.,  {Nayakshin} S.,   {Rosotti} G.,  2022, \mn@doi
  [\mnras] {10.1093/mnras/stac1774}, \href
  {https://ui.adsabs.harvard.edu/abs/2022MNRAS.515.3113E} {515, 3113}

\bibitem[\protect\citeauthoryear{{Facchini}, {Pinilla}, {van Dishoeck}  \& {de
  Juan Ovelar}}{{Facchini} et~al.}{2018}]{facchini2018}
{Facchini} S.,  {Pinilla} P.,  {van Dishoeck} E.~F.,   {de Juan Ovelar} M.,
  2018, \mn@doi [\aap] {10.1051/0004-6361/201731390}, \href
  {https://ui.adsabs.harvard.edu/abs/2018A&A...612A.104F} {612, A104}

\bibitem[\protect\citeauthoryear{{Flaherty} et~al.,}{{Flaherty}
  et~al.}{2017}]{Flaherty2017}
{Flaherty} K.~M.,  et~al., 2017, \mn@doi [\apj] {10.3847/1538-4357/aa79f9},
  \href {https://ui.adsabs.harvard.edu/abs/2017ApJ...843..150F} {843, 150}

\bibitem[\protect\citeauthoryear{{Flaherty} et~al.,}{{Flaherty}
  et~al.}{2020}]{Flaherty2020}
{Flaherty} K.,  et~al., 2020, \mn@doi [\apj] {10.3847/1538-4357/ab8cc5}, \href
  {https://ui.adsabs.harvard.edu/abs/2020ApJ...895..109F} {895, 109}

\bibitem[\protect\citeauthoryear{{Follette} et~al.,}{{Follette}
  et~al.}{2022}]{Follette2022}
{Follette} K.~B.,  et~al., 2022, \mn@doi [arXiv e-prints]
  {10.48550/arXiv.2211.02109}, \href
  {https://ui.adsabs.harvard.edu/abs/2022arXiv221102109F} {p. arXiv:2211.02109}

\bibitem[\protect\citeauthoryear{{Fu}, {Li}, {Lubow}  \& {Li}}{{Fu}
  et~al.}{2014}]{Fu2014}
{Fu} W.,  {Li} H.,  {Lubow} S.,   {Li} S.,  2014, \mn@doi [\apjl]
  {10.1088/2041-8205/788/2/L41}, \href
  {https://ui.adsabs.harvard.edu/abs/2014ApJ...788L..41F} {788, L41}

\bibitem[\protect\citeauthoryear{{Goldreich} \& {Tremaine}}{{Goldreich} \&
  {Tremaine}}{1980}]{Goldreich_Tremaine_1980}
{Goldreich} P.,  {Tremaine} S.,  1980, \mn@doi [\apj] {10.1086/158356}, \href
  {http://adsabs.harvard.edu/abs/1980ApJ...241..425G} {241, 425}

\bibitem[\protect\citeauthoryear{{Hammer}, {Lin}, {Kratter}  \&
  {Pinilla}}{{Hammer} et~al.}{2021}]{HammerEtal2021}
{Hammer} M.,  {Lin} M.-K.,  {Kratter} K.~M.,   {Pinilla} P.,  2021, \mn@doi
  [\mnras] {10.1093/mnras/stab1079}, \href
  {https://ui.adsabs.harvard.edu/abs/2021MNRAS.504.3963H} {504, 3963}

\bibitem[\protect\citeauthoryear{{Hasegawa}, {Okuzumi}, {Flock}  \&
  {Turner}}{{Hasegawa} et~al.}{2017}]{Hasegawa2017}
{Hasegawa} Y.,  {Okuzumi} S.,  {Flock} M.,   {Turner} N.~J.,  2017, \mn@doi
  [\apj] {10.3847/1538-4357/aa7d55}, \href
  {https://ui.adsabs.harvard.edu/abs/2017ApJ...845...31H} {845, 31}

\bibitem[\protect\citeauthoryear{{Huang} et~al.,}{{Huang}
  et~al.}{2018}]{2018HuangDSHARP2}
{Huang} J.,  et~al., 2018, \mn@doi [\apjl] {10.3847/2041-8213/aaf740}, \href
  {http://adsabs.harvard.edu/abs/2018ApJ...869L..42H} {869, L42}

\bibitem[\protect\citeauthoryear{{Huang}, {Li}, {Isella}, {Miranda}, {Li}  \&
  {Ji}}{{Huang} et~al.}{2020}]{HuangEtal2020p}
{Huang} P.,  {Li} H.,  {Isella} A.,  {Miranda} R.,  {Li} S.,   {Ji} J.,  2020,
  \mn@doi [\apj] {10.3847/1538-4357/ab8199}, \href
  {https://ui.adsabs.harvard.edu/abs/2020ApJ...893...89H} {893, 89}

\bibitem[\protect\citeauthoryear{{Hu{\'e}lamo} et~al.,}{{Hu{\'e}lamo}
  et~al.}{2022}]{Huelamo2022}
{Hu{\'e}lamo} N.,  et~al., 2022, \mn@doi [\aap] {10.1051/0004-6361/202243918},
  \href {https://ui.adsabs.harvard.edu/abs/2022A&A...668A.138H} {668, A138}

\bibitem[\protect\citeauthoryear{{Ilee}, {Hall}, {Walsh}, {Jim{\'e}nez-Serra},
  {Pinte}, {Terry}, {Bourke}  \& {Hoare}}{{Ilee} et~al.}{2020}]{Ilee2020-SKA}
{Ilee} J.~D.,  {Hall} C.,  {Walsh} C.,  {Jim{\'e}nez-Serra} I.,  {Pinte} C.,
  {Terry} J.,  {Bourke} T.~L.,   {Hoare} M.,  2020, \mn@doi [\mnras]
  {10.1093/mnras/staa2699}, \href
  {https://ui.adsabs.harvard.edu/abs/2020MNRAS.498.5116I} {498, 5116}

\bibitem[\protect\citeauthoryear{{Jiang} \& {Ormel}}{{Jiang} \&
  {Ormel}}{2021}]{JiangOrmel2021}
{Jiang} H.,  {Ormel} C.~W.,  2021, \mn@doi [\mnras] {10.1093/mnras/stab1278},
  \href {https://ui.adsabs.harvard.edu/abs/2021MNRAS.505.1162J} {505, 1162}

\bibitem[\protect\citeauthoryear{{Jorquera} et~al.,}{{Jorquera}
  et~al.}{2021}]{Jorquera2021}
{Jorquera} S.,  et~al., 2021, \mn@doi [\aj] {10.3847/1538-3881/abd40d}, \href
  {https://ui.adsabs.harvard.edu/abs/2021AJ....161..146J} {161, 146}

\bibitem[\protect\citeauthoryear{{Keppler} et~al.,}{{Keppler}
  et~al.}{2019}]{KepplerEtal2019}
{Keppler} M.,  et~al., 2019, \mn@doi [\aap] {10.1051/0004-6361/201935034},
  \href {https://ui.adsabs.harvard.edu/abs/2019A&amp;A...625A.118K} {625, A118}

\bibitem[\protect\citeauthoryear{{Kimmig}, {Dullemond}  \& {Kley}}{{Kimmig}
  et~al.}{2020}]{Kimmig-2020}
{Kimmig} C.~N.,  {Dullemond} C.~P.,   {Kley} W.,  2020, \mn@doi [\aap]
  {10.1051/0004-6361/201936412}, \href
  {https://ui.adsabs.harvard.edu/abs/2020A&A...633A...4K} {633, A4}

\bibitem[\protect\citeauthoryear{{Lehmann} \& {Lin}}{{Lehmann} \&
  {Lin}}{2022}]{LehmannLin2022}
{Lehmann} M.,  {Lin} M.~K.,  2022, \mn@doi [\aap]
  {10.1051/0004-6361/202142378}, \href
  {https://ui.adsabs.harvard.edu/abs/2022A&amp;A...658A.156L} {658, A156}

\bibitem[\protect\citeauthoryear{{Lesur}}{{Lesur}}{2021}]{Lesur2021}
{Lesur} G. R.~J.,  2021, \mn@doi [\aap] {10.1051/0004-6361/202040109}, \href
  {https://ui.adsabs.harvard.edu/abs/2021A&A...650A..35L} {650, A35}

\bibitem[\protect\citeauthoryear{{Li}, {Colgate}, {Wendroff}  \& {Liska}}{{Li}
  et~al.}{2001}]{Li2001}
{Li} H.,  {Colgate} S.~A.,  {Wendroff} B.,   {Liska} R.,  2001, \mn@doi [\apj]
  {10.1086/320241}, \href
  {https://ui.adsabs.harvard.edu/abs/2001ApJ...551..874L} {551, 874}

\bibitem[\protect\citeauthoryear{{Lin} \& {Papaloizou}}{{Lin} \&
  {Papaloizou}}{1986}]{Lin_Papaloizou1986a}
{Lin} D.~N.~C.,  {Papaloizou} J.,  1986, \mn@doi [\apj] {10.1086/164426}, \href
  {https://ui.adsabs.harvard.edu/abs/1986ApJ...307..395L} {307, 395}

\bibitem[\protect\citeauthoryear{{Lin} \& {Papaloizou}}{{Lin} \&
  {Papaloizou}}{1993}]{Lin_Papaloizou_1993}
{Lin} D.~N.~C.,  {Papaloizou} J.~C.~B.,  1993, in {Levy} E.~H.,  {Lunine}
  J.~I.,  eds, Protostars and Planets III. pp 749--835

\bibitem[\protect\citeauthoryear{{Lodato}, {Scardoni}, {Manara}  \&
  {Testi}}{{Lodato} et~al.}{2017}]{Lodato2017}
{Lodato} G.,  {Scardoni} C.~E.,  {Manara} C.~F.,   {Testi} L.,  2017, \mn@doi
  [\mnras] {10.1093/mnras/stx2273}, \href
  {https://ui.adsabs.harvard.edu/abs/2017MNRAS.472.4700L} {472, 4700}

\bibitem[\protect\citeauthoryear{{Lodato} et~al.,}{{Lodato}
  et~al.}{2019}]{lodato2019}
{Lodato} G.,  et~al., 2019, \mn@doi [\mnras] {10.1093/mnras/stz913}, \href
  {https://ui.adsabs.harvard.edu/abs/2019MNRAS.486..453L} {486, 453}

\bibitem[\protect\citeauthoryear{{Long} et~al.,}{{Long}
  et~al.}{2018}]{long2018}
{Long} F.,  et~al., 2018, \mn@doi [\apj] {10.3847/1538-4357/aae8e1}, \href
  {https://ui.adsabs.harvard.edu/abs/2018ApJ...869...17L} {869, 17}

\bibitem[\protect\citeauthoryear{{Long} et~al.,}{{Long}
  et~al.}{2019}]{Long2019}
{Long} F.,  et~al., 2019, \mn@doi [\apj] {10.3847/1538-4357/ab2d2d}, \href
  {https://ui.adsabs.harvard.edu/abs/2019ApJ...882...49L} {882, 49}

\bibitem[\protect\citeauthoryear{{Long} et~al.,}{{Long}
  et~al.}{2022}]{LongEtal2022}
{Long} F.,  et~al., 2022, \mn@doi [\apj] {10.3847/1538-4357/ac634e}, \href
  {https://ui.adsabs.harvard.edu/abs/2022ApJ...931....6L} {931, 6}

\bibitem[\protect\citeauthoryear{{Lovelace}, {Li}, {Colgate}  \&
  {Nelson}}{{Lovelace} et~al.}{1999}]{RWI-1999}
{Lovelace} R.~V.~E.,  {Li} H.,  {Colgate} S.~A.,   {Nelson} A.~F.,  1999,
  \mn@doi [\apj] {10.1086/306900}, \href
  {https://ui.adsabs.harvard.edu/abs/1999ApJ...513..805L} {513, 805}

\bibitem[\protect\citeauthoryear{{Mac{\'\i}as}, {Anglada}, {Osorio},
  {Torrelles}, {Carrasco-Gonz{\'a}lez}, {G{\'o}mez}, {Rodr{\'\i}guez}  \&
  {Sierra}}{{Mac{\'\i}as} et~al.}{2017}]{MaciasEtal2017}
{Mac{\'\i}as} E.,  {Anglada} G.,  {Osorio} M.,  {Torrelles} J.~M.,
  {Carrasco-Gonz{\'a}lez} C.,  {G{\'o}mez} J.~F.,  {Rodr{\'\i}guez} L.~F.,
  {Sierra} A.,  2017, \mn@doi [\apj] {10.3847/1538-4357/aa6620}, \href
  {https://ui.adsabs.harvard.edu/abs/2017ApJ...838...97M} {838, 97}

\bibitem[\protect\citeauthoryear{{Marino}, {Casassus}, {Perez}, {Lyra},
  {Roman}, {Avenhaus}, {Wright}  \& {Maddison}}{{Marino}
  et~al.}{2015}]{MarinoEtal2015}
{Marino} S.,  {Casassus} S.,  {Perez} S.,  {Lyra} W.,  {Roman} P.~E.,
  {Avenhaus} H.,  {Wright} C.~M.,   {Maddison} S.~T.,  2015, \mn@doi [\apj]
  {10.1088/0004-637X/813/1/76}, \href
  {https://ui.adsabs.harvard.edu/abs/2015ApJ...813...76M} {813, 76}

\bibitem[\protect\citeauthoryear{{Mathis}, {Rumpl}  \& {Nordsieck}}{{Mathis}
  et~al.}{1977}]{MRN}
{Mathis} J.~S.,  {Rumpl} W.,   {Nordsieck} K.~H.,  1977, \mn@doi [\apj]
  {10.1086/155591}, \href
  {https://ui.adsabs.harvard.edu/abs/1977ApJ...217..425M} {217, 425}

\bibitem[\protect\citeauthoryear{{Miotello}, {Kamp}, {Birnstiel}, {Cleeves}  \&
  {Kataoka}}{{Miotello} et~al.}{2022}]{Miotello2022}
{Miotello} A.,  {Kamp} I.,  {Birnstiel} T.,  {Cleeves} L.~I.,   {Kataoka} A.,
  2022, arXiv e-prints, \href
  {https://ui.adsabs.harvard.edu/abs/2022arXiv220309818M} {p. arXiv:2203.09818}

\bibitem[\protect\citeauthoryear{{Miranda}, {Li}, {Li}  \& {Jin}}{{Miranda}
  et~al.}{2017}]{Miranda2017}
{Miranda} R.,  {Li} H.,  {Li} S.,   {Jin} S.,  2017, \mn@doi [\apj]
  {10.3847/1538-4357/835/2/118}, \href
  {https://ui.adsabs.harvard.edu/abs/2017ApJ...835..118M} {835, 118}

\bibitem[\protect\citeauthoryear{{Nakagawa}, {Sekiya}  \& {Hayashi}}{{Nakagawa}
  et~al.}{1986}]{NakagawaEtal1986}
{Nakagawa} Y.,  {Sekiya} M.,   {Hayashi} C.,  1986, \mn@doi [\icarus]
  {10.1016/0019-1035(86)90121-1}, \href
  {https://ui.adsabs.harvard.edu/abs/1986Icar...67..375N} {67, 375}

\bibitem[\protect\citeauthoryear{{Ono}, {Muto}, {Takeuchi}  \& {Nomura}}{{Ono}
  et~al.}{2016}]{Ono2016}
{Ono} T.,  {Muto} T.,  {Takeuchi} T.,   {Nomura} H.,  2016, \mn@doi [\apj]
  {10.3847/0004-637X/823/2/84}, \href
  {https://ui.adsabs.harvard.edu/abs/2016ApJ...823...84O} {823, 84}

\bibitem[\protect\citeauthoryear{{Paardekooper} \& {Mellema}}{{Paardekooper} \&
  {Mellema}}{2006}]{paardekooper2006}
{Paardekooper} S.~J.,  {Mellema} G.,  2006, \mn@doi [\aap]
  {10.1051/0004-6361:20054449}, \href
  {https://ui.adsabs.harvard.edu/abs/2006A&A...453.1129P} {453, 1129}

\bibitem[\protect\citeauthoryear{{Paardekooper}, {Dong}, {Duffell}, {Fung},
  {Masset}, {Ogilvie}  \& {Tanaka}}{{Paardekooper}
  et~al.}{2022}]{PaardekooperEtal2022}
{Paardekooper} S.-J.,  {Dong} R.,  {Duffell} P.,  {Fung} J.,  {Masset} F.~S.,
  {Ogilvie} G.,   {Tanaka} H.,  2022, arXiv e-prints, \href
  {https://ui.adsabs.harvard.edu/abs/2022arXiv220309595P} {p. arXiv:2203.09595}

\bibitem[\protect\citeauthoryear{{P{\'e}rez} et~al.,}{{P{\'e}rez}
  et~al.}{2018}]{PerezEtal2018}
{P{\'e}rez} L.~M.,  et~al., 2018, \mn@doi [\apjl] {10.3847/2041-8213/aaf745},
  \href {https://ui.adsabs.harvard.edu/abs/2018ApJ...869L..50P} {869, L50}

\bibitem[\protect\citeauthoryear{{P{\'e}rez}, {Casassus}, {Baruteau}, {Dong},
  {Hales}  \& {Cieza}}{{P{\'e}rez} et~al.}{2019}]{PerezEtal2019}
{P{\'e}rez} S.,  {Casassus} S.,  {Baruteau} C.,  {Dong} R.,  {Hales} A.,
  {Cieza} L.,  2019, \mn@doi [\aj] {10.3847/1538-3881/ab1f88}, \href
  {https://ui.adsabs.harvard.edu/abs/2019AJ....158...15P} {158, 15}

\bibitem[\protect\citeauthoryear{{Pinilla} et~al.,}{{Pinilla}
  et~al.}{2018}]{PinillaEtal2018}
{Pinilla} P.,  et~al., 2018, \mn@doi [\apj] {10.3847/1538-4357/aae824}, \href
  {https://ui.adsabs.harvard.edu/abs/2018ApJ...868...85P} {868, 85}

\bibitem[\protect\citeauthoryear{{Pinte}, {Dent}, {M{\'e}nard}, {Hales},
  {Hill}, {Cortes}  \& {de Gregorio-Monsalvo}}{{Pinte}
  et~al.}{2016}]{Pinte2016}
{Pinte} C.,  {Dent} W.~R.~F.,  {M{\'e}nard} F.,  {Hales} A.,  {Hill} T.,
  {Cortes} P.,   {de Gregorio-Monsalvo} I.,  2016, \mn@doi [\apj]
  {10.3847/0004-637X/816/1/25}, \href
  {https://ui.adsabs.harvard.edu/abs/2016ApJ...816...25P} {816, 25}

\bibitem[\protect\citeauthoryear{{Poblete} et~al.,}{{Poblete}
  et~al.}{2022}]{PobleteEtal2022}
{Poblete} P.~P.,  et~al., 2022, \mn@doi [\mnras] {10.1093/mnras/stab3474},
  \href {https://ui.adsabs.harvard.edu/abs/2022MNRAS.510..205P} {510, 205}

\bibitem[\protect\citeauthoryear{{Ricci}, {Liu}, {Isella}  \& {Li}}{{Ricci}
  et~al.}{2018}]{Ricci2018-ngVLA}
{Ricci} L.,  {Liu} S.-F.,  {Isella} A.,   {Li} H.,  2018, \mn@doi [\apj]
  {10.3847/1538-4357/aaa546}, \href
  {https://ui.adsabs.harvard.edu/abs/2018ApJ...853..110R} {853, 110}

\bibitem[\protect\citeauthoryear{{Riols} \& {Lesur}}{{Riols} \&
  {Lesur}}{2019}]{Riols2019}
{Riols} A.,  {Lesur} G.,  2019, \mn@doi [\aap] {10.1051/0004-6361/201834813},
  \href {https://ui.adsabs.harvard.edu/abs/2019A&A...625A.108R} {625, A108}

\bibitem[\protect\citeauthoryear{{Rometsch}, {Ziampras}, {Kley}  \&
  {B{\'e}thune}}{{Rometsch} et~al.}{2021}]{RometschEtal2021}
{Rometsch} T.,  {Ziampras} A.,  {Kley} W.,   {B{\'e}thune} W.,  2021, \mn@doi
  [\aap] {10.1051/0004-6361/202142105}, \href
  {https://ui.adsabs.harvard.edu/abs/2021A&amp;A...656A.130R} {656, A130}

\bibitem[\protect\citeauthoryear{{Rosotti}}{{Rosotti}}{2023}]{Giovanni-2023}
{Rosotti} G.~P.,  2023, \mn@doi [\nar] {10.1016/j.newar.2023.101674}, \href
  {https://ui.adsabs.harvard.edu/abs/2023NewAR..9601674R} {96, 101674}

\bibitem[\protect\citeauthoryear{{Rosotti}, {Juhasz}, {Booth}  \&
  {Clarke}}{{Rosotti} et~al.}{2016}]{Rosotti2016}
{Rosotti} G.~P.,  {Juhasz} A.,  {Booth} R.~A.,   {Clarke} C.~J.,  2016, \mn@doi
  [\mnras] {10.1093/mnras/stw691}, \href
  {http://adsabs.harvard.edu/abs/2016MNRAS.459.2790R} {459, 2790}

\bibitem[\protect\citeauthoryear{{Rosotti}, {Teague}, {Dullemond}, {Booth}  \&
  {Clarke}}{{Rosotti} et~al.}{2020}]{Rosotti2020}
{Rosotti} G.~P.,  {Teague} R.,  {Dullemond} C.,  {Booth} R.~A.,   {Clarke}
  C.~J.,  2020, \mn@doi [\mnras] {10.1093/mnras/staa1170}, \href
  {https://ui.adsabs.harvard.edu/abs/2020MNRAS.495..173R} {495, 173}

\bibitem[\protect\citeauthoryear{{Shakura} \& {Sunyaev}}{{Shakura} \&
  {Sunyaev}}{1973}]{Shakura-Sunyaev73}
{Shakura} N.~I.,  {Sunyaev} R.~A.,  1973, \aap, \href
  {http://cdsads.u-strasbg.fr/cgi-bin/nph-bib_query?bibcode=1973A%26A....24..337S&db_key=AST}
  {24, 337}

\bibitem[\protect\citeauthoryear{{Stadler} et~al.,}{{Stadler}
  et~al.}{2023}]{StadlerEtal2023}
{Stadler} J.,  et~al., 2023, \mn@doi [\aap] {10.1051/0004-6361/202245381},
  \href {https://ui.adsabs.harvard.edu/abs/2023A&amp;A...670L...1S} {670, L1}

\bibitem[\protect\citeauthoryear{{Stammler}, {Lichtenberg},
  {Dr{\k{a}}{\.z}kowska}  \& {Birnstiel}}{{Stammler}
  et~al.}{2023}]{StammlerEtal2023}
{Stammler} S.~M.,  {Lichtenberg} T.,  {Dr{\k{a}}{\.z}kowska} J.,   {Birnstiel}
  T.,  2023, \mn@doi [\aap] {10.1051/0004-6361/202245512}, \href
  {https://ui.adsabs.harvard.edu/abs/2023A&amp;A...670L...5S} {670, L5}

\bibitem[\protect\citeauthoryear{{Suriano}, {Li}, {Krasnopolsky}, {Suzuki}  \&
  {Shang}}{{Suriano} et~al.}{2019}]{Suriano2019}
{Suriano} S.~S.,  {Li} Z.-Y.,  {Krasnopolsky} R.,  {Suzuki} T.~K.,   {Shang}
  H.,  2019, \mn@doi [\mnras] {10.1093/mnras/sty3502}, \href
  {https://ui.adsabs.harvard.edu/abs/2019MNRAS.484..107S} {484, 107}

\bibitem[\protect\citeauthoryear{{Suzuki}, {Ogihara}, {Morbidelli}, {Crida}  \&
  {Guillot}}{{Suzuki} et~al.}{2016}]{Suzuki2016}
{Suzuki} T.~K.,  {Ogihara} M.,  {Morbidelli} A.,  {Crida} A.,   {Guillot} T.,
  2016, \mn@doi [\aap] {10.1051/0004-6361/201628955}, \href
  {https://ui.adsabs.harvard.edu/abs/2016A&A...596A..74S} {596, A74}

\bibitem[\protect\citeauthoryear{{Tabone}, {Rosotti}, {Lodato}, {Armitage},
  {Cridland}  \& {van Dishoeck}}{{Tabone} et~al.}{2022a}]{Tabone2022-2}
{Tabone} B.,  {Rosotti} G.~P.,  {Lodato} G.,  {Armitage} P.~J.,  {Cridland}
  A.~J.,   {van Dishoeck} E.~F.,  2022a, \mn@doi [\mnras]
  {10.1093/mnrasl/slab124}, \href
  {https://ui.adsabs.harvard.edu/abs/2022MNRAS.512L..74T} {512, L74}

\bibitem[\protect\citeauthoryear{{Tabone}, {Rosotti}, {Cridland}, {Armitage}
  \& {Lodato}}{{Tabone} et~al.}{2022b}]{Tabone22}
{Tabone} B.,  {Rosotti} G.~P.,  {Cridland} A.~J.,  {Armitage} P.~J.,   {Lodato}
  G.,  2022b, \mn@doi [\mnras] {10.1093/mnras/stab3442}, \href
  {https://ui.adsabs.harvard.edu/abs/2022MNRAS.512.2290T} {512, 2290}

\bibitem[\protect\citeauthoryear{{Toci}, {Lodato}, {Livio}, {Rosotti}  \&
  {Trapman}}{{Toci} et~al.}{2023}]{TociEtal2023}
{Toci} C.,  {Lodato} G.,  {Livio} F.~G.,  {Rosotti} G.,   {Trapman} L.,  2023,
  \mn@doi [\mnras] {10.1093/mnrasl/slac137}, \href
  {https://ui.adsabs.harvard.edu/abs/2023MNRAS.518L..69T} {518, L69}

\bibitem[\protect\citeauthoryear{{Trapman}, {Rosotti}, {Bosman}, {Hogerheijde}
  \& {van Dishoeck}}{{Trapman} et~al.}{2020}]{TrapmanEtal2020}
{Trapman} L.,  {Rosotti} G.,  {Bosman} A.~D.,  {Hogerheijde} M.~R.,   {van
  Dishoeck} E.~F.,  2020, \mn@doi [\aap] {10.1051/0004-6361/202037673}, \href
  {https://ui.adsabs.harvard.edu/abs/2020A&amp;A...640A...5T} {640, A5}

\bibitem[\protect\citeauthoryear{{Trapman}, {Tabone}, {Rosotti}  \&
  {Zhang}}{{Trapman} et~al.}{2022}]{TrapmanEtal2022a}
{Trapman} L.,  {Tabone} B.,  {Rosotti} G.,   {Zhang} K.,  2022, \mn@doi [\apj]
  {10.3847/1538-4357/ac3ed5}, \href
  {https://ui.adsabs.harvard.edu/abs/2022ApJ...926...61T} {926, 61}

\bibitem[\protect\citeauthoryear{{Villenave} et~al.,}{{Villenave}
  et~al.}{2020}]{Villenave2020}
{Villenave} M.,  et~al., 2020, \mn@doi [\aap] {10.1051/0004-6361/202038087},
  \href {https://ui.adsabs.harvard.edu/abs/2020A&A...642A.164V} {642, A164}

\bibitem[\protect\citeauthoryear{{Weber}, {P{\'e}rez}, {Ben{\'\i}tez-Llambay},
  {Gressel}, {Casassus}  \& {Krapp}}{{Weber}
  et~al.}{2019}]{Weber2019-dust-diffusion}
{Weber} P.,  {P{\'e}rez} S.,  {Ben{\'\i}tez-Llambay} P.,  {Gressel} O.,
  {Casassus} S.,   {Krapp} L.,  2019, \mn@doi [\apj]
  {10.3847/1538-4357/ab412f}, \href
  {https://ui.adsabs.harvard.edu/abs/2019ApJ...884..178W} {884, 178}

\bibitem[\protect\citeauthoryear{{Wu} \& {Liu}}{{Wu} \&
  {Liu}}{2023}]{Wu2023-SKA}
{Wu} Y.,  {Liu} S.-F.,  2023, in prep

\bibitem[\protect\citeauthoryear{{Youdin} \& {Lithwick}}{{Youdin} \&
  {Lithwick}}{2007}]{Youdin-Lithwick}
{Youdin} A.~N.,  {Lithwick} Y.,  2007, \mn@doi [\icarus]
  {10.1016/j.icarus.2007.07.012}, \href
  {https://ui.adsabs.harvard.edu/abs/2007Icar..192..588Y} {192, 588}

\bibitem[\protect\citeauthoryear{{Zagaria}, {Rosotti}, {Clarke}  \&
  {Tabone}}{{Zagaria} et~al.}{2022}]{ZagariaEtal2022}
{Zagaria} F.,  {Rosotti} G.~P.,  {Clarke} C.~J.,   {Tabone} B.,  2022, \mn@doi
  [\mnras] {10.1093/mnras/stac1461}, \href
  {https://ui.adsabs.harvard.edu/abs/2022MNRAS.514.1088Z} {514, 1088}

\bibitem[\protect\citeauthoryear{{Zhang} et~al.,}{{Zhang}
  et~al.}{2018}]{2018ZhangDSHARP}
{Zhang} S.,  et~al., 2018, \mn@doi [\apjl] {10.3847/2041-8213/aaf744}, \href
  {https://ui.adsabs.harvard.edu/abs/2018ApJ...869L..47Z} {869, L47}

\bibitem[\protect\citeauthoryear{{Zhou}, {Deng}, {Chen}  \& {Lin}}{{Zhou}
  et~al.}{2022}]{Zhou2022}
{Zhou} T.,  {Deng} H.-P.,  {Chen} Y.-X.,   {Lin} D. N.~C.,  2022, \mn@doi
  [\apj] {10.3847/1538-4357/ac9bf6}, \href
  {https://ui.adsabs.harvard.edu/abs/2022ApJ...940..117Z} {940, 117}

\bibitem[\protect\citeauthoryear{{Zhu} \& {Wu}}{{Zhu} \&
  {Wu}}{2018}]{ZhuWu2018}
{Zhu} W.,  {Wu} Y.,  2018, \mn@doi [\aj] {10.3847/1538-3881/aad22a}, \href
  {https://ui.adsabs.harvard.edu/abs/2018AJ....156...92Z} {156, 92}

\bibitem[\protect\citeauthoryear{{Zhu}, {Nelson}, {Hartmann}, {Espaillat}  \&
  {Calvet}}{{Zhu} et~al.}{2011}]{Zhuetal2011}
{Zhu} Z.,  {Nelson} R.~P.,  {Hartmann} L.,  {Espaillat} C.,   {Calvet} N.,
  2011, \mn@doi [\apj] {10.1088/0004-637X/729/1/47}, \href
  {https://ui.adsabs.harvard.edu/abs/2011ApJ...729...47Z} {729, 47}

\bibitem[\protect\citeauthoryear{{Zhu}, {Stone}  \& {Bai}}{{Zhu}
  et~al.}{2015}]{Zhu-2015}
{Zhu} Z.,  {Stone} J.~M.,   {Bai} X.-N.,  2015, \mn@doi [\apj]
  {10.1088/0004-637X/801/2/81}, \href
  {https://ui.adsabs.harvard.edu/abs/2015ApJ...801...81Z} {801, 81}

\bibitem[\protect\citeauthoryear{{van der Marel} et~al.,}{{van der Marel}
  et~al.}{2013}]{vanderMarelEtal2013a}
{van der Marel} N.,  et~al., 2013, \mn@doi [Science] {10.1126/science.1236770},
  \href {https://ui.adsabs.harvard.edu/abs/2013Sci...340.1199V} {340, 1199}

\bibitem[\protect\citeauthoryear{{van der Marel}, {Cazzoletti}, {Pinilla}  \&
  {Garufi}}{{van der Marel} et~al.}{2016}]{vanderMarelEtal2016i}
{van der Marel} N.,  {Cazzoletti} P.,  {Pinilla} P.,   {Garufi} A.,  2016,
  \mn@doi [\apj] {10.3847/0004-637X/832/2/178}, \href
  {https://ui.adsabs.harvard.edu/abs/2016ApJ...832..178V} {832, 178}

\bibitem[\protect\citeauthoryear{{van der Marel} et~al.,}{{van der Marel}
  et~al.}{2021}]{vanderMarelEtal2021a}
{van der Marel} N.,  et~al., 2021, \mn@doi [\aj] {10.3847/1538-3881/abc3ba},
  \href {https://ui.adsabs.harvard.edu/abs/2021AJ....161...33V} {161, 33}

\makeatother
\end{thebibliography}





\bsp	
\label{lastpage}
\end{CJK*}
\end{document}